\documentclass{emulateapj}
\usepackage{colordvi}
\usepackage{epsfig}
\usepackage{graphicx}
\usepackage{amssymb}
\newcommand{\Ha}{\hbox{{\rm H}\kern 0.1em$\alpha$}}
\newcommand{\Hb}{\hbox{{\rm H}\kern 0.1em$\beta$}}
\newcommand{\MgII}{\hbox{{\rm Mg}\kern 0.1em{\sc ii}}}
\newcommand{\CIV}{\hbox{{\rm C}\kern 0.1em{\sc iv}}}
\newcommand{\NeV}{\hbox{[{\rm Ne}\kern 0.1em{\sc v}]}}
\newcommand{\OII}{\hbox{[{\rm O}\kern 0.1em{\sc ii}]}}
\newcommand{\NeIII}{\hbox{[{\rm Ne}\kern 0.1em{\sc iii}]}}
\newcommand{\OIII}{\hbox{[{\rm O}\kern 0.1em{\sc iii}]}}
\newcommand{\NII}{\hbox{[{\rm N}\kern 0.1em{\sc ii}]}}
\newcommand{\SII}{\hbox{[{\rm S}\kern 0.1em{\sc ii}]}}
\newcommand{\sfrtot}{SFR$_{\mathrm{IR+UV}}$}
\newcommand{\suny}{M$_{\odot}$~y$^{-1}$}
\newcommand{\lssfr}{log(sSFR)}
\newcommand{\lmass}{log(M/M$_{\odot}$)}
\newcommand{\mdyn}{M$_{\rm{dyn}}$}
\newcommand{\mgas}{M$_{\rm{gas}}$}
\newcommand{\mstar}{M$_{\star}$}
\newcommand{\mdm}{M$_{\rm{DM}}$}
\newcommand{\lratio}{log(M$_{\star}$/M$_{\rm{dyn}}$)}
\newcommand{\lgas}{M$_{\rm{gas}}$/M$_{\rm{dyn}}$}

\newcommand{\sigint}{$\sigma_{\rm{int}}$}

\begin{document}

\title{Keck-I MOSFIRE spectroscopy of compact star-forming galaxies at \lowercase{z}$\gtrsim$2: High velocity dispersions in progenitors of compact quiescent galaxies}

\author{Guillermo Barro\altaffilmark{1},
Jonathan R.~Trump\altaffilmark{2,3},
David C.~Koo\altaffilmark{1}, 
Avishai Dekel\altaffilmark{4},
Susan A.~Kassin\altaffilmark{5},
Dale D.~Kocevski\altaffilmark{6}, 
Sandra M. Faber\altaffilmark{1},
Arjen van der Wel\altaffilmark{7},
Yicheng Guo\altaffilmark{1},
Pablo G.~P\'{e}rez-Gonz\'{a}lez\altaffilmark{8,9},
Elisa Toloba\altaffilmark{1},
Jerome J. Fang\altaffilmark{1}, 
Camilla Pacifici\altaffilmark{10},
Raymond Simons\altaffilmark{11},
Randy D. Campbell\altaffilmark{12},
Daniel Ceverino\altaffilmark{13},
Steven L. Finkelstein\altaffilmark{14},
Bob Goodrich\altaffilmark{12},
Marc Kassis\altaffilmark{12},
Anton M. Koekemoer\altaffilmark{5},
Nicholas P. Konidaris\altaffilmark{15},
Rachael C. Livermore\altaffilmark{14},
James E. Lyke\altaffilmark{12},
Bahram Mobasher\altaffilmark{16},
Hooshang Nayyeri\altaffilmark{16},
Michael Peth\altaffilmark{11},
Joel R. Primack\altaffilmark{17},
Luca Rizzi\altaffilmark{12},
Rachel S. Somerville\altaffilmark{18},
Gregory D. Wirth\altaffilmark{12},
Adi Zolotov\altaffilmark{4},
}

\altaffiltext{1}{University of California, Santa Cruz}
\altaffiltext{2}{Pennsylvania State University}
\altaffiltext{3}{Hubble Fellow}
\altaffiltext{4}{The Hebrew University}
\altaffiltext{5}{Space Telescope Science Institute}
\altaffiltext{6}{University of Kentucky}
\altaffiltext{7}{Max-Planck-Institut f\"{u}r Astronomie}
\altaffiltext{8}{Universidad Complutense de Madrid}
\altaffiltext{9}{Steward Observatory, University of Arizona}
\altaffiltext{10}{Yonsei University Observatory}
\altaffiltext{11}{Johns Hopkins University}
\altaffiltext{12}{W. M. Keck Observatory}
\altaffiltext{13}{Universidad Autonoma de Madrid}
\altaffiltext{14}{The University of Texas at Austin}
\altaffiltext{15}{California Institute of Technology}
\altaffiltext{16}{University of California, Riverside}
\altaffiltext{17}{Santa Cruz Institute for Particle Physics}
\altaffiltext{18}{Rutgers University}

\slugcomment{Submitted to the Astrophysical Journal}

%\slugcomment{Last edited: \today} \date{Submitted: \today}
%\pagerange{\pageref{firstpage}--\pageref{lastpage}} \pubyear{2008}
%\maketitle
\label{firstpage}
\begin{abstract} 
We present Keck-I MOSFIRE near-infrared spectroscopy for a sample of
13 compact star-forming galaxies (SFGs) at redshift $2\leq z \leq2.5$
with star formation rates of SFR~$\sim$~100~\suny~and masses of
\lmass~$\sim10.8$. Their high integrated gas velocity dispersions of
\sigint$=230^{+40}_{-30}$~km s$^{-1}$, as measured from emission lines
of \Ha~and \OIII, and the resultant \mstar$-$\sigint~relation and
\mstar$-$\mdyn~all match well to those of compact quiescent galaxies
at $z\sim2$, as measured from stellar absorption lines. Since
\lratio~$=-0.06\pm0.2$~dex, these compact SFGs appear to be
dynamically relaxed and more evolved, i.e., more depleted in gas and
dark matter ($<$13$^{+17}_{-13}$\%) than their non-compact SFG
counterparts at the same epoch. Without infusion of external gas,
depletion timescales are short, less than $\sim$300~Myr. This
discovery adds another link to our new dynamical chain of evidence
that compact SFGs at $z\gtrsim2$ are already losing gas to become the
immediate progenitors of compact quiescent galaxies by $z\sim2$.

\end{abstract}
\keywords{galaxies: starburst --- galaxies: photometry --- galaxies:
  high-redshift}

\section{Introduction}\label{intro}

\begin{figure*}[t]
\centering
\includegraphics[width=18cm,angle=0.]{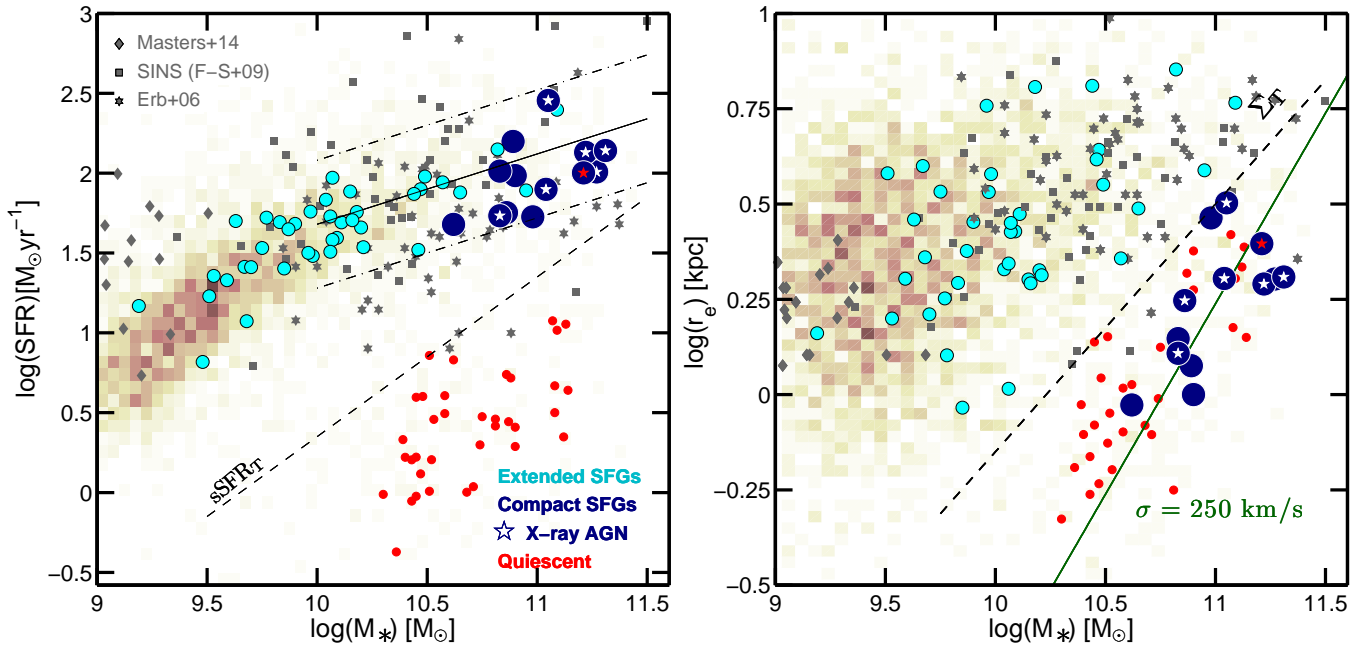}
\caption{\label{selectiondiag} {\it Left panel:} SFR--mass diagram for
  all galaxies in the parent galaxy catalog in GOODS-S and -N at
  $2<z<3$. The checkered gray scale illustrates the location of the
  star-forming main sequence. The solid and dashed-dotted lines depict
  the best fit and 1~$\sigma$ scatter to the massive end (\lmass$>$10
  ) of the main-sequence. The 13 compact and 67 extended SFGs observed
  with NIR spectroscopy are shown in blue and cyan colors,
  respectively. X-ray detected galaxies are indicated with a white
  star symbol. The red star depicts the galaxy in common with
  \citet[][see \S~\ref{kinprop}]{belli14b}. The dashed line
  illustrates the selection threshold in sSFR
  (\lssfr$<-1$~Gyr$^{-1}$). The red markers show the quiescent
  population . {\it Right panel:} mass--size distribution for the same
  galaxies in the left panel. The dashed line illustrates the
  compactness threshold
  ($\Sigma_{1.5}=10.4~M_{\odot}$kpc$^{-1.5}$). The green line shows
  the expected location of galaxies with constant velocity dispersion
  $\sigma=250$~km s$^{-1}$, as inferred from Equation~3, assuming
  \mdyn=\mstar. These panels illustrate our selection criteria for
  compact SFGs aimed at identifying {\it normal} main-sequence SFGs
  following a more compact mass-size relation, similar to that of the
  quiescent population.}
\end{figure*}

The formation scenario for the first massive quiescent galaxies is
still unclear. While observations report a rapid increase in the
number density of massive galaxies with suppressed star formation
rates (SFRs) since $z\sim3$ (\citealt{brammer11};
\citealt{whitaker11}; \citealt{muzzin13smf}), other surprising
results, such as their remarkably small ($\sim$1~kpc scale) sizes
(\citealt{trujillo07}; \citealt{buitrago08}; \citealt{dokku08};
\citealt{cassata11}; \citealt{szo12}) and their large velocity
dispersions (\citealt{dokku08}; \citealt{toft12}; \citealt{newman13};
\citealt{vandesande13}; \citealt{bezanson13}; \citealt{belli14a}) in
comparison with local galaxies of the same stellar mass, pose a puzzle
as to which galaxies are their immediate star-forming
progenitors. Given that galaxy structure and kinematics appear to be
more robust and stable properties than luminosity or SFR (e.g.,
\citealt{franx08}; \citealt{wake12}; \citealt{bell12};
\citealt{cheung12}), it is unlikely that the typical massive SFGs at
$z\sim2$, which consist mostly of extended disks (\citealt{wuyts11b};
\citealt{buitrago13}) with irregular, sometimes clumpy, morphologies
and have high rotational velocities
(\citealt{elmegreen04,elmegreen05}; \citealt{genzel08,genzel12};
\citealt{law09,law12a}; \citealt{fs09,fs11}; \citealt{wuyts12};
\citealt{swinbank12c,swinbank12b}; \citealt{guo12}) are the
progenitors of compact quiescent galaxies. Instead, it is more
plausible that their precursors are similarly compact SFGs. These
compact SFGs probably formed in strongly dissipative gas-rich
processes, such as mergers (\citealt{hopkins06}; \citealt{naab07};
\citealt{wuyts10}) or accretion-driven disk instabilities
(\citealt{dekel09a,dekel13b}; \citealt{ceverino10}) which contracts
the galaxy, producing a compact, dispersion-dominated, remnant
(\citealt{dekel13b}).

A crucial step forward in determining whether the latter scenario is
at work is the recent discovery of a population of massive
(\lmass~$>10.3$), {\it compact} dusty SFGs at $z\gtrsim2$
(\citealt{wuyts11b}; \citealt{barro13}; \citealt{patel13};
\citealt{stefanon13}).  In \citet{barro13,barro14} we showed that
compact SFGs have dust-obscured spectral energy distributions (SEDs)
characterized by bright IR fluxes ($\sim$70\% and 30\% are detected by
{\it Spitzer} and {\it Herschel} in the far IR) that, nonetheless,
translate into seemingly {\it normal} SFR~$\sim100-200$~\suny,
different from those of extreme submillimeter (sub-mm) galaxies
(\citealt{bothwell13}; \citealt{toft14}).  Structurally, they present
centrally-concentrated, spheroidal morphologies and high S\'ersic
indices consistent with those of compact quiescent galaxies, and they
follow a similar mass-size relation.  In addition, we found that the
observed number density of compact SFGs can reproduce the build up of
the compact quiescent population since $z\sim3$, if they quench
star-formation in a few 10$^{8}$yr. This led us to propose an
evolutionary picture in which compact SFGs are formed from larger SFGs
as a result of gas-rich processes (mergers or disk-instabilities) that
induce a compact starburst which quench on dynamical timescales fading
into a compact quiescent galaxy. So far the similarities between
compact SFGs and quiescent galaxies are already very
encouraging. However, they are based on photometric and structural
properties and thus have yet to be verified from kinematic data to
confirm the connection.

This paper presents near-IR (NIR) spectroscopic follow-up of a sample
of massive compact SFGs at $2\leq z \leq 3$ presented in Barro et
al. (2013,14) to measure their kinematic properties and compare them
against those of compact quiescent galaxies to test whether they
support the picture of a rapid fading into the red sequence.  We also
estimate stellar and dynamical masses for compact SFGs to infer their
gas fractions and gas depletion timescales, and we analyze their
stacked spectra for signs of outflowing gas. While several other
surveys have already presented emission line measurements and
kinematic properties for SFGs at $z\gtrsim1.5$ (\citealt{erb06};
\citealt{law07}; \citealt{law09}; \citealt{fs09}; \citealt{epinat12};
\citealt{snewman13}; \citealt{masters14}; \citealt{buitrago14};
\citealt{rwilliams14}), this is the first observational effort to
target specifically massive, yet small, compact SFGs, which may be
notoriously missing in those surveys.

Throughout the paper, we adopt a flat cosmology with $\Omega_{M}$=0.3,
$\Omega_{\Lambda}$=0.7 and H$_{0}=70$~km~s$^{-1}$~Mpc$^{-1}$ and we
quote magnitudes in the AB system.

\section{DATA \& SAMPLE SELECTION}\label{data}

Our targets are compact SFGs at $z\gtrsim2$ to be measured for
kinematic properties using their emission lines. These galaxies, first
identified in \citet{wuyts11b} and \citet{barro13}, have been proposed
to be the immediate precursors of compact quiescent galaxies at
$z\sim2$ as they share structural properties while spanning a range in
SFRs, from main-sequence to almost quenched \citep{barro14}.

\subsection{Photometric data, stellar properties and SFRs}\label{stellarprop}

We select galaxies from the CANDELS (\citealt{candelsgro};
\citealt{candelskoe}) WFC3/F160W ($H$-band) multi-wavelength catalogs
in GOODS-S, GOODS-N and COSMOS (\citealt{guo13}; Barro et al. in
prep.; Nayyeri et al. in prep.). The galaxy spectral energy
distributions (SEDs) include extensive multi-band data ranging from
the UV to the near-IR.  We also include complementary mid-IR
photometry in {\it Spitzer}/MIPS 24 and 70~$\mu$m (30~$\mu$Jy and
1~mJy, $5\sigma$) from \citet{pg08b}, and far-IR from the
GOODS-Herschel (\citealt{elbaz11}) and PEP (\citealt{magnelli13})
surveys. For each galaxy, we fit photometric redshifts using EAzY
\citep{eazy} and calculate stellar masses using FAST \citep{fast}
assuming \cite{bc03} stellar population synthesis models, a
\cite{chabrier} initial mass function (IMF) and the \cite{calzetti}
dust extinction law with attenuations ranging between $0<A_{V}<4$. We
assume an exponentially declining star formation history with
timescale $\tau$ and age $t$. Following \citet{wuyts11a} we impose a
soft constraint on the minimum $e$-folding time (log~$\tau>8.5$) to
obtain better agreement between different SFR indicators (see
below). Age are allowed to vary over the range 10~Myr~$<t<t_{H}$, where
$t_{H}$ is the age of the universe at the given redshift.

We follow the method of the SFR-{\it ladder} as described in
\citet{wuyts11a} to obtain consistent SFRs over a broad dynamic
range. In brief, for galaxies detected at mid-to-far IR wavelengths
(i.e., {\it Spitzer}/MIPS and {\it Herschel}/PACS) we compute the
total SFR by adding the unobscured and obscured star formation, traced
by the UV and IR emission, respectively, following \citet[][see also
  \citealt{bell05}]{ken98}.

\begin{equation}
SFR_{\mathrm{UV+IR}}=1.09\times10^{-10}(L_{\mathrm{IR}}+3.3L_{2800})[M_{\odot}/yr]
\end{equation}

\noindent
where $L_{\mathrm{IR}}$ is the total IR luminosity
($L_{\mathrm{IR}}\equiv L(8-1000~\mu m)$) derived from the fit to {\it
  Spitzer} and {\it Herschel} data, and $L_{2800}$ is estimated from
the best fitting SED template. The normalization factor corresponds to
a \citet{chabrier} IMF. For galaxies undetected in the IR
(\sfrtot~$\lesssim$~30~\suny) we correct SFR$_{\mathrm{UV}}$ for
extinction using the attenuation derived from the best-fit SED
model. This method has been shown to provide consistent SFR estimates
down to very low sSFR levels (log(sSFR)$>-1$~Gyr$^{-1}$;
\citealt{wuyts11a}; \citealt{fumagalli13}; \citealt{utomo14}). In
practice, we used \sfrtot~ for all compact SFGs described in the next
section, as they are all detected in MIPS 24~$\mu$m and $\sim30\%$ in
PACS.

X-ray source identifications and total luminosities
($L$$_{\mathrm{X}}\equiv L_{0.5-8\mathrm{kev}}$) were computed for the
sources identified in the {\it Chandra} 4~Ms and 2~Ms catalogs in
GOODS-S \citep{chandra4m} and GOODS-N \citep{chandra2m}, respectively.

The shape of the two-dimensional surface brightness profiles measured
from the {\it HST}/WFC3 F160W image were modeled using GALFIT
\citep{galfit}. The effective (half-light) radius and the S\'ersic
index, $n$ are determined using a single component fit. Position
dependent point spread functions (PSFs) are created and processed with
TinyTim \citep{tinytim} to replicate the conditions of the observed
data when fitting light profiles. The method and the catalog of
morphological properties are fully described in \citet{vdw12}. The
stellar and star-formation properties have been previously used in
several other papers (\citealt{wuyts11b}; \citealt{barro13,barro14};
\citealt{trump13}; Guo et al. in prep.).

\begin{figure*}[t]
\centering
\includegraphics[width=8.9cm,angle=0.]{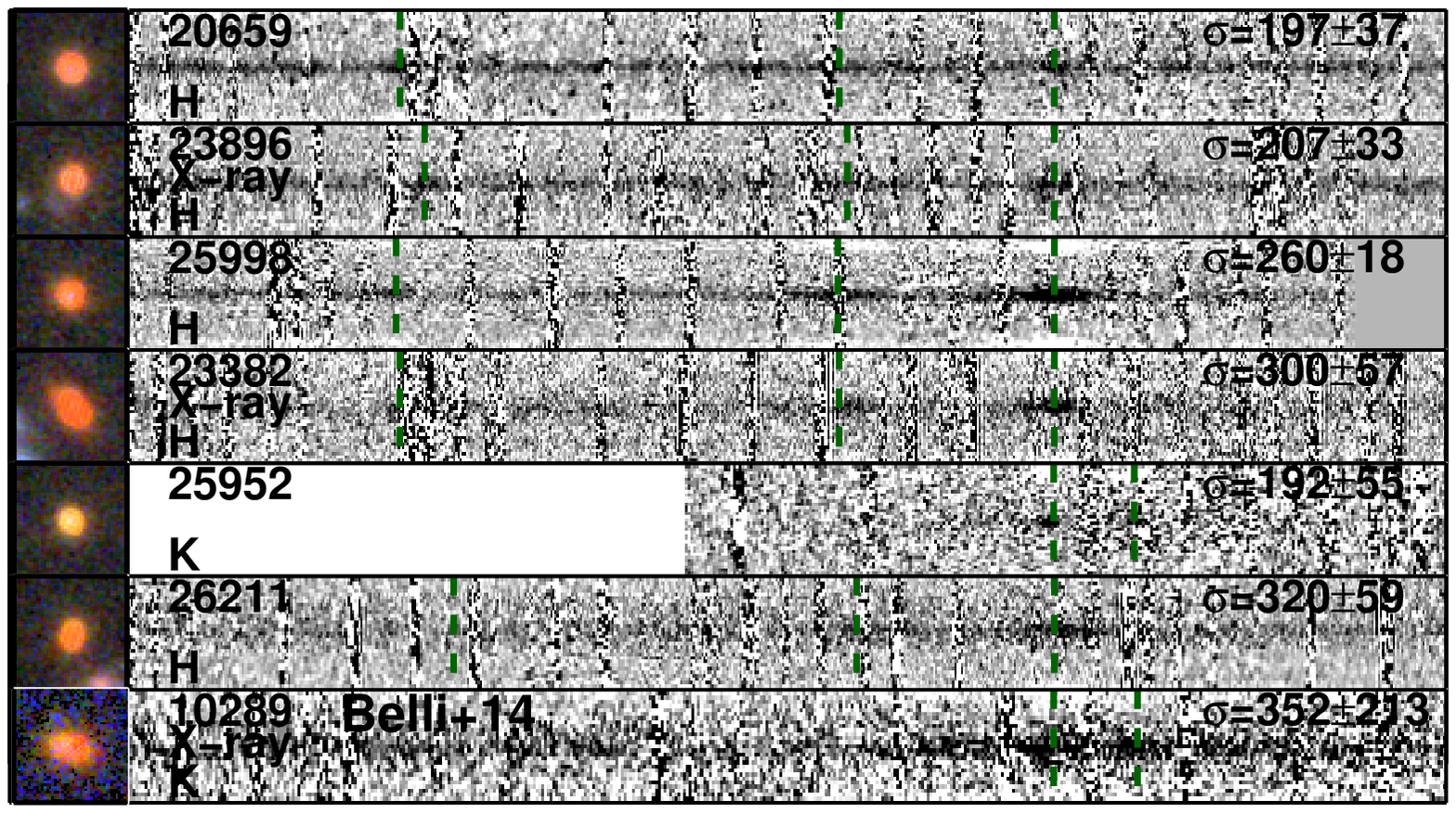}
\includegraphics[width=8.9cm,angle=0.]{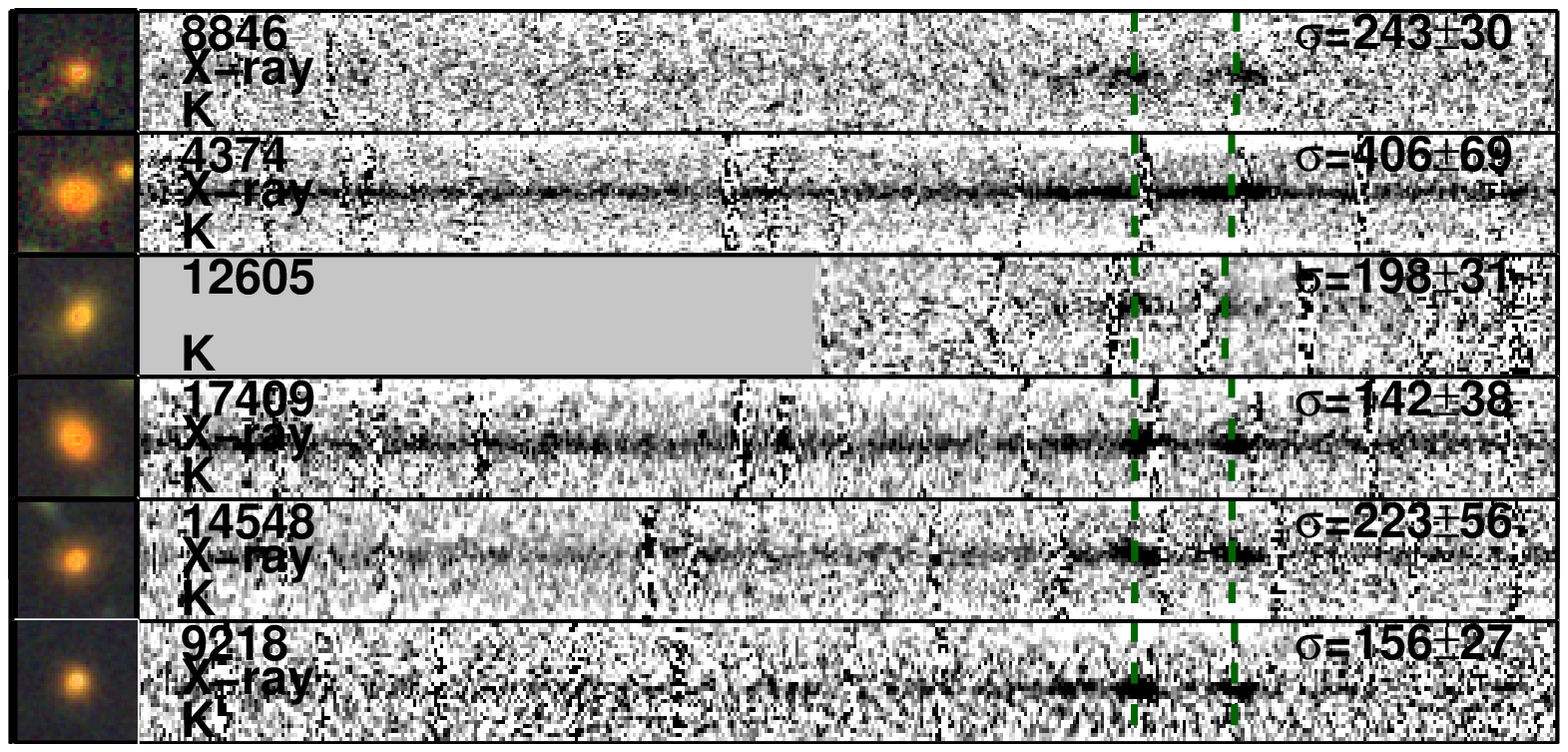}
\caption{\label{2dspectra} 2D MOSFIRE $H-$ and $K-$ spectra of the 13
  compact SFGs jointly with their 3''$\times$3'' ACS/WFC3 $zJH$ color
  composite postage stamps. Compact SFGs have strikingly red
  rest-frame UV-optical colors due to their high dust obscuration. The
  green dashed lines indicate the most prominent emission lines,
  either \Ha~and/or \OIII~and \Hb~with increasing wavelength to the
  right (see also Figure~\ref{spectra} for more details). Compact SFGs
  are barely resolved in the spatial direction. Only 17409 and 14548
  show a hint of resolved kinematics. The measured \sigint~and the
  X-ray detected galaxies are indicated in the text.}
\end{figure*}

\begin{figure*}
\centering
\includegraphics[width=18cm,angle=0.]{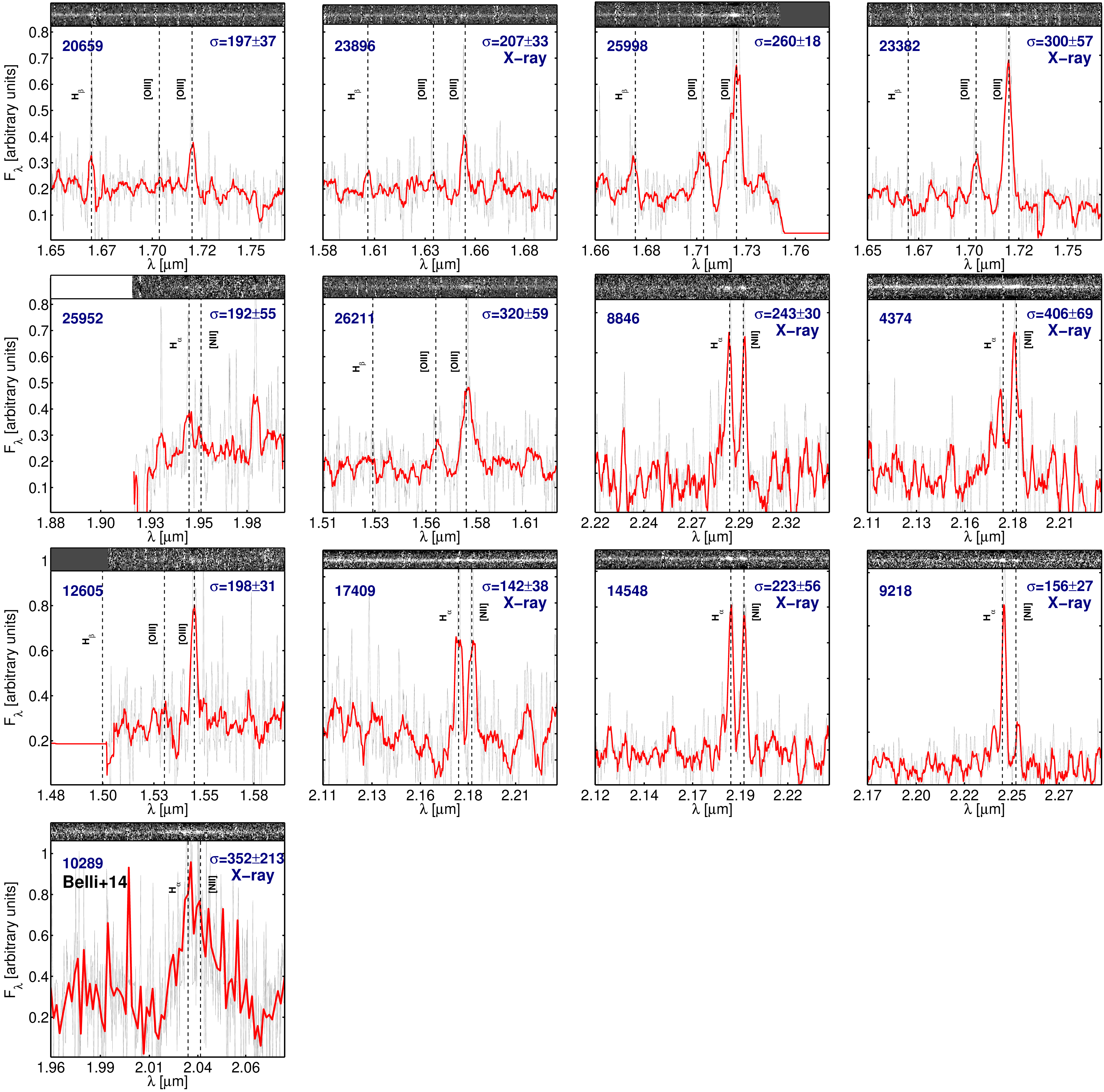}
\caption{\label{spectra} 1D MOSFIRE $H-$ and $K-$ spectra of the 13
  compact SFGs. Their ID, \sigint, and detection in the X-rays is
  indicated. The gray line shows the collapsed 1D spectra extracted
  from the 2D spectra shown in Figure~\ref{2dspectra}. The red line
  shows the binned 1D spectra. The dashed lines indicate the most
  prominent emission lines in the observed spectral range. The last
  row shows the spectra of the galaxy in common with the sample of
  quiescent galaxies of \citet{belli14b}.}
\end{figure*}

\subsection{Selection of compact and extended SFGs}

We select compact SFGs following the method described in
\citet{barro13,barro14}. Briefly, we require galaxies to be
star-forming by having \lssfr$>-9.5$~Gyr$^{-1}$, roughly the mass
doubling time (1/$t_{H}$) at $z\sim2$. We define compactness using a
threshold in {\it pseudo} stellar mass surface density of
$\Sigma_{1.5}=$log$(M/r^{1.5}_{\mathrm{e}})>10.4~M_{\odot}$kpc$^{-1.5}$.
Figure~\ref{selectiondiag} illustrates this selection showing the
location of the 13 compact SFGs (blue circles) observed with MOSFIRE
overlaid in the SFR-mass and mass-size diagrams for galaxies more
massive than \lmass$>9$ at $2<z<3$ in the CANDELS GOODS-N and -S
catalogs. The compactness criterion selects SFGs that follow a
mass-size relation similar to that of the quiescent population. As
shown in \citet[][see also \citealt{cassata13};
  \citealt{vdw14}]{newman12}, quiescent galaxies follow a tight
mass-size relation with a slope, $\alpha\sim1.5$, that remains
constant with redshift, and a mass-normalized radius that evolves with
(1+z)$^{0.025}$.

In order to discuss the properties of compact SFGs in the general
context of SFGs at $z\sim2$, Figure~\ref{selectiondiag} shows a sample
of 67 extended SFGs ($\Sigma_{1.5}<10.4$; cyan circles), also observed
in our MOSFIRE survey, and other SFGs at $z\gtrsim1.4-2.5$ drawn from
\citet{erb06}, \citet{fs09}, \citet{maseda13} and
\citet{masters14}. It is clear from the figure that, while compact
SFGs lie in the same locus of the SFR-mass diagram as other SFGs, they
occupy a distinct region of the mass-size diagram under-represented in
previous surveys of SFGs. For simplicity, in the remainder of the
paper we use the term extended SFGs to describe both our extended SFGs
and those from the reference samples, which also have
$\Sigma_{1.5}<10.4$.

\section{Mosfire NIR spectroscopy}

We conducted NIR spectroscopic observations in GOODS-S using the new
multi-object spectrograph MOSFIRE (\citealt{mosfire1,mosfire2}) on the
Keck-I telescope. The data were taken in 4 runs on September $14-15$,
October 10 2012, December $25-28$ 2013 and January $1-2$ 2014.  We
observed a total of 8 masks in $H-$band ($1.46<\lambda<1.81~\mu$m) and
3 masks in $K-$band ($1.93<\lambda<2.41~\mu$m), with exposure times
ranging between $50-120$ minutes, and $40-100$ minutes,
respectively. Overall, the weather conditions were excellent with
seeing $\sim0\farcs4-0\farcs6$ and good transparency in all the masks
except 2 in the $K-$band, for which the seeing was poor ($\gtrsim$1'')
and there were partial clouds. We used the same observational
configuration for all masks: 2-point dithers separated by $1\farcs5$
and slit widths of $0\farcs7$. The instrumental resolution of MOSFIRE
with $0\farcs7$ slit widths is approximately $R=3200$ ($\sim$5~\AA~per
resolution element). 2D spectra were reduced, sky subtracted,
wavelength calibrated, and one-dimensionaly extracted using the public
MOSFIRE data reduction pipeline (Konidaris in prep.). Redshifts were
found using the Specpro software \citep{specpro}. We also include in
this paper SFGs observed as a part of the TKRS2 survey, which observed
the GOODS-N field in the $J$, $H$ and $K$-bands in 3 runs on December
27 and Januray 14 2012, and May 3 2013 (see Wirth in prep. for details
on the overall target selection and observing strategy), and 1
$K$-band mask in COSMOS observed for 60min in March 14 2014 (PI:
Finkelstein). The data were reduced following the same procedure as in
our main program.

Overall each MOSFIRE mask includes 25 galaxies out of which only a
small fraction are discussed in this paper. Our primary targets are 13
compact SFGs and 67 extended SFGs at $z=1.8-2.4$. The latter were
selected to study the excitation properties of AGNs and the kinematic
properties of clumpy SFGs (see \citealt{trump13} and Guo et al. 2014
in prep. for more details on these galaxies). Here we used those
galaxies mostly for comparison purposes to illustrate the differences
relative to compact SFGs.

\section{Dynamical properties of compact SFGs}

\subsection{Kinematic measurements}\label{kinmes}

Here we assume that the emission-line velocity dispersions are
primarily due to the motion of the gas in the gravitational potential
of the galaxy. Nevertheless, in the following sections, we also
discuss the possible caveats in this assumption due to the presence of
an AGN or to the effects of turbulence and outflows. The integrated
velocity dispersion, \sigint, can be determined for all galaxies
because it requires only a measurement of the width of the emission
line. It is therefore the most straightforward and useful kinematic
quantity. Note that, even in the best seeing conditions
($\sim0\farcs4$), compact SFGs are barely resolved in the spatial
direction (Figure~\ref{2dspectra}, see also \S~\ref{outflows}). Thus
\sigint~could represent either the intrinsic $\sigma$ of the galaxy,
the collapsed $v_{\rm{rot}}$ of a small rotating disk or a combination
of the two.

We determine the one-dimensional velocity dispersion by fitting a
Gaussian profile to each emission line, measuring its FWHM, and
subtracting the instrumental broadening in quadrature from the FWHM
(following \citealt{weiner06} and \citealt{kassin07}). The instrumental
broadening is measured from the widths of skylines and is 2.7\AA~ (in
the observed frame). The velocity dispersion is then the corrected
FWHM divided by 2.355. The whole sample of compact SFGs consists of 13
galaxies with velocity dispersions ranging from \sigint$=140-400$~km
s$^{-1}$ and an average value of \sigint$=230^{+40}_{-30}$~km
s$^{-1}$. For the 67 extended galaxies, the velocity dispersions are
typically lower, ranging from \sigint$=66-150$~km s$^{-1}$. Many of
the extended galaxies are spatially resolved at the average seeing of
the observations, and thus their rotational and dispersion components
can be estimated from the 2D spectra. These results will be presented
in Guo et al. (2014 in prep.) and Simons et al. (2014 in prep.). Here we
adopt the \sigint~values to provide a homogeneous kinematic
measurement for all galaxies.

\subsection{Kinematics of AGN hosts}\label{agn}

More than half of the compact SFGs are detected in the X-rays (7/13),
and have large X-ray luminosities L$_{\rm{X}}>10^{43}$~erg/s which
suggest the presence of an AGN. This could cause a potential bias on
the measured linewidths, as AGN-ionized gas in the "narrow line
region" (NLR) is typically dominated by emission nearer the center of
a galaxy.  Even though this gas is usually dominated by the
gravitational potential of the galaxy rather than the AGN itself, it
has broader velocities than gas at the larger effective radius.

The 7 X-ray detected galaxies, however, do not have systematically
broader velocity dispersions than the X-ray undetected galaxies.
Instead, their median \sigint$=200$~km s$^{-1}$, fully consistent with
that of the non X-ray detected compact SFGs.  This is perhaps
unsurprising, as their X-ray luminosities are roughly at the detection
threshold of the X-ray data at $z\sim2-3$, rather than at large,
QSO-like, values (L$_{\rm{X}}>10^{44}$~erg/s).  We also find no
evidence for UV or IR excess in the IRAC bands ($[8.0]/[3.6] \lesssim
1.3$; \citealt{donley07,donley12}) as a result of the AGN emission
and, for the 5 Herschel/PACS detected galaxies, the median far-IR
colors, $[160]/[24]=40$, are fully consistent with star-formation
(\citealt{kirkpatric13}; \citealt{barro14}). The only exception is
galaxy 10289 which appears to have an elevated 8~$\mu$m flux
indicative of hot dust emission near the AGN (we discuss this galaxy
in the next section). Overall, these tests suggest that the AGN does
not have a strong effect on the SED or the line widths inferred from
it.

The ratios between partially ionized forbidden lines and recombination
lines also suggest that compact SFGs are not strongly AGN-dominated.
While we only have both \OIII/\Hb~and \NII/\Ha~for one galaxy to
calculate the full BPT diagram \citep{bpt}, we do have \NII/\Ha~alone
for all but 2 galaxies.  If we combine this \NII/\Ha~with a
conservative, but elevated, \OIII/\Hb~$\sim 5$, similar to values
reported in recent papers (\citealt{trump11b}; \citealt{holden14};
\citealt{steidel14}), these galaxies would lie predominantly in the
mixed AGN - star forming region with 0.2$\leq$\NII/\Ha$\leq$1 values
(see Table 1). Only 2 galaxies present \NII/\Ha~ratios significantly
larger than one (25998 and 4374), and both of them are clearly
detected by {\it Herschel} indicating on-going star formation. In
fact, 4374, which presents the largest \sigint~of the sample, is a
sub-mm galaxy (see e.g., \citealt{laird10}; \citealt{michalowski12})

Interestingly, even if the emission lines are partially fueled by the
AGN rather than star-formation, these galaxies are so compact that the
AGN narrow line region (NLR) probably corresponds to the whole galaxy.
For example, \citet{bennert02} show that the size of the NLR
correlates with the \OIII~luminosity, reaching sizes $r_{\rm{NLR}} >
1$~kpc for $L_{\rm{[OIII]}} \sim 10^{43}$~erg/s, which is consistent
with the expected values for these strongly star-forming galaxies with
enhanced \OIII/\Hb~ratios. Therefore their linewidths are still good
tracers of the potential well.

\begin{figure}[t]
\centering \includegraphics[width=9cm,angle=0.]{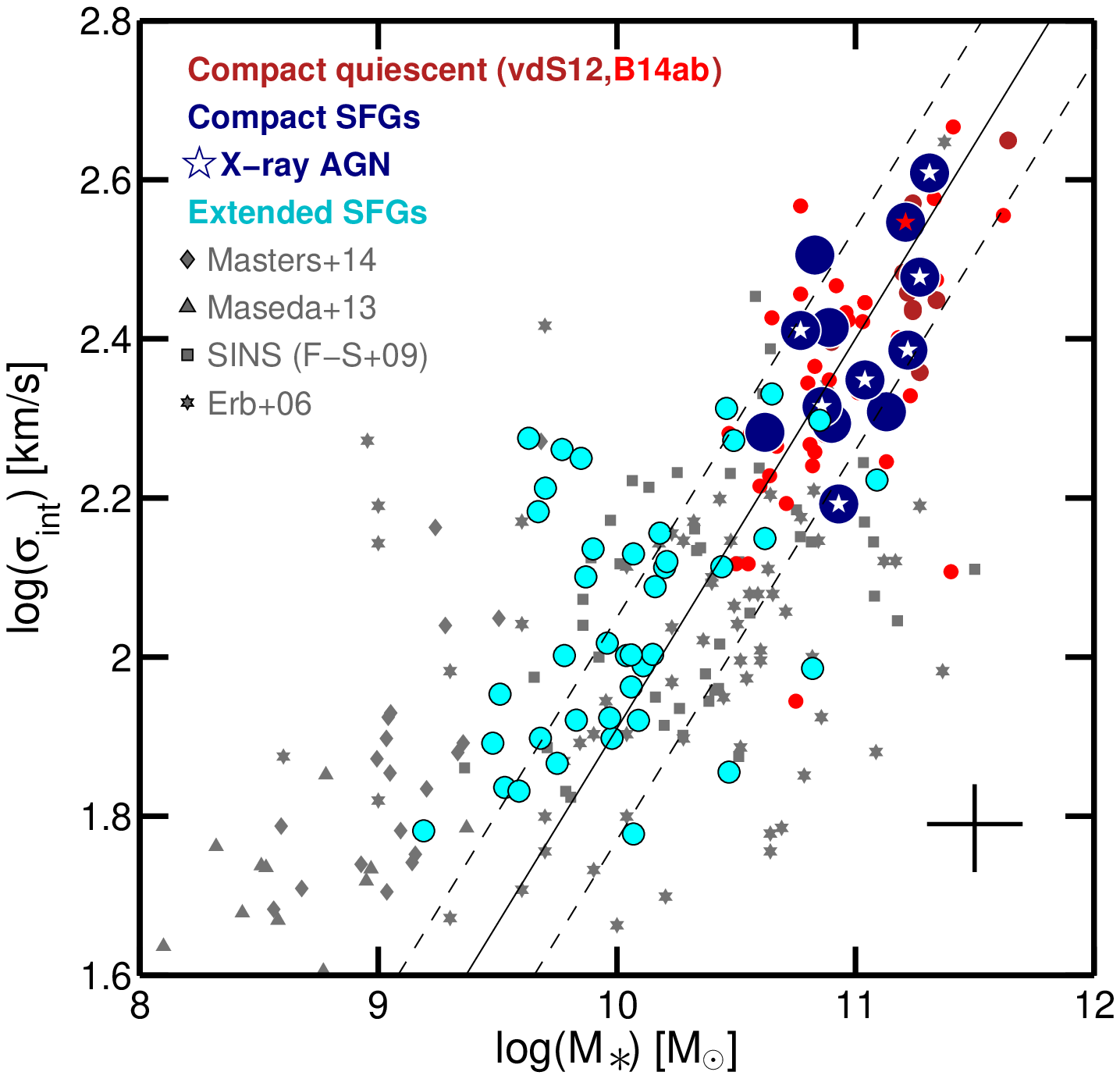}
\caption{\label{faberjackson} Distribution of \sigint~vs. \mstar~for
  compact (blue) and extended (cyan, and gray for other references)
  SFGs with MOSFIRE spectra. AGN hosts are indicated with white
  stars. The red star depicts the galaxy in common with
  \citet{belli14b}. The error bars in the bottom-right corner indicate
  the average uncertainty in \sigint~and \mstar~for compact and
  extended SFGs. Compact SFGs and quiescent galaxies occupy the same
  region of the diagram presenting a similar
  \mstar$-$\sigint~correlation (solid line) with a tight scatter
  ($\Delta$log(\sigint)$=$0.14~dex; dashed lines) over $\sim$1~dex in
  stellar mass.}
\end{figure}

\subsection{Kinematics of Compact SFGs and Quiescent galaxies}\label{kinprop}

In order to study whether the similarities between compact SFGs and
quiescent galaxies extend beyond their stellar masses and structural
properties, Figure~\ref{faberjackson} compares the integrated
kinematics vs. mass for both populations. Quite remarkably, compact
SFGs occupy the same region of the diagram, following a similar
\mstar$-$\sigint~relation over more than 1~dex in stellar mass, as
that of a compilation of compact quiescent galaxies at
$1.5\leq z \lesssim2$ drawn from \citet{vandesande13} and
\citet{belli14a,belli14b}. The \sigint~values in quiescent galaxies
are measured from absorption lines and therefore use stars instead of
gas as a tracer of the gravitational potential.  This could
potentially introduce systematic offsets if gas and stars present
different motions due to shocks or turbulence, or if they are located
in different regions of the galaxy. However, the striking similarity
in the distribution of both populations suggest that compact SFGs are
in fact kinematically relaxed (i.e., \sigint~traces the gravitational
potential) and their integrated properties match those of quiescent
galaxies.

Obtaining more conclusive evidence of the matching kinematic
properties of compact SFGs and quiescent requires comparing emission
and absorption line measurements of the same galaxies. This is
observationally very challenging because: a) it requires long
integrations to detect absorption lines at $z\sim2$, and b) absorption
surveys target only quiescent galaxies without emission
lines. However, recent observations by \citet{belli14b} present
absorption line kinematics for one compact SFGs in our sample (10289;
marked with a red star in all Figures), and the dispersion values are
consistent within the 1$\sigma$ errors, $\sigma_{\rm{abs}}=312\pm65$
km s$^{-1}$ vs. $\sigma_{\rm{int}}=357\pm213$ km s$^{-1}$. The
uncertainty in the emission line measurement is larger due to the
shorter integration time and the likely dusty nature of the
galaxy. However, the \Ha~broadening is apparent in both the 2D- and
1D- spectra, and the peak of the emission line lies clearly between
the strong sky lines (Figures~\ref{2dspectra} and \ref{spectra}). The
galaxy presents a IR-excess at $\lambda_{\rm{rest}}=3$~$\mu$m, likely
coming from the AGN, that suggests that the IR-based SFR is
overestimated.  Nevertheless, the SED-fit also suggests an elevated
dust extinction ($A_{\rm{V}}=2.0$) and a UV-based SFR$=$80~\suny above
the sSFR selection threshold for SFGs.

Figure~\ref{faberjackson} also reveals that extended SFGs tend to
deviate from the \mstar$-$\sigint~relation of compact galaxies having
larger \sigint~at a given stellar mass, particularly at
\lmass$\lesssim$10.  The black line shows the best-fit
\mstar$-$\sigint~relation for compact SFG and quiescent galaxies
(log(\sigint)$=$(1.91$\pm$0.07)+(0.49$\pm$0.04)(\lmass-10);
$\Delta$log(\sigint)$=$0.14~dex), which presents a steeper slope than
the local Faber-Jackson relation (\citealt{faber76};
\citealt{gallazzi06}), as also noted in \citealt{belli14a}.  Extended
(disk-dominated) galaxies, are often represented in the
\mstar$-$v$_{\rm{rot}}$, Tully-Fisher relation
\citep{tully77}. However, using a value that accounts for both
rotation and velocity dispersion, such as \sigint~, they can be shown
together with other galaxies in a more fundamental relation that is a
tracer of the total dynamical mass of the galaxy (e.g.;
\citealt{weiner06}; \citealt{kassin07}; \citealt{cappellari13};
\citealt{courteau14}). In that context, the trend in low-mass extended
SFGs could indicate, that \sigint~is not only gravitational (i.e.,
there are turbulent motions), or perhaps that \sigint~overestimates
the intrinsic contributions from rotation and dispersion.
Alternatively, extended SFGs could indeed follow a shallower
\mstar$-$\sigint~relation because they have different dynamical masses
than the more massive compact galaxies, i.e., they may have a larger
contribution of dark matter or gas mass to the gravitational
potential, thus increasing the velocity dispersion beyond the expected
value if \mdyn$\sim$\mstar.

\begin{figure*}[t]
\centering
\includegraphics[width=18cm,angle=0.]{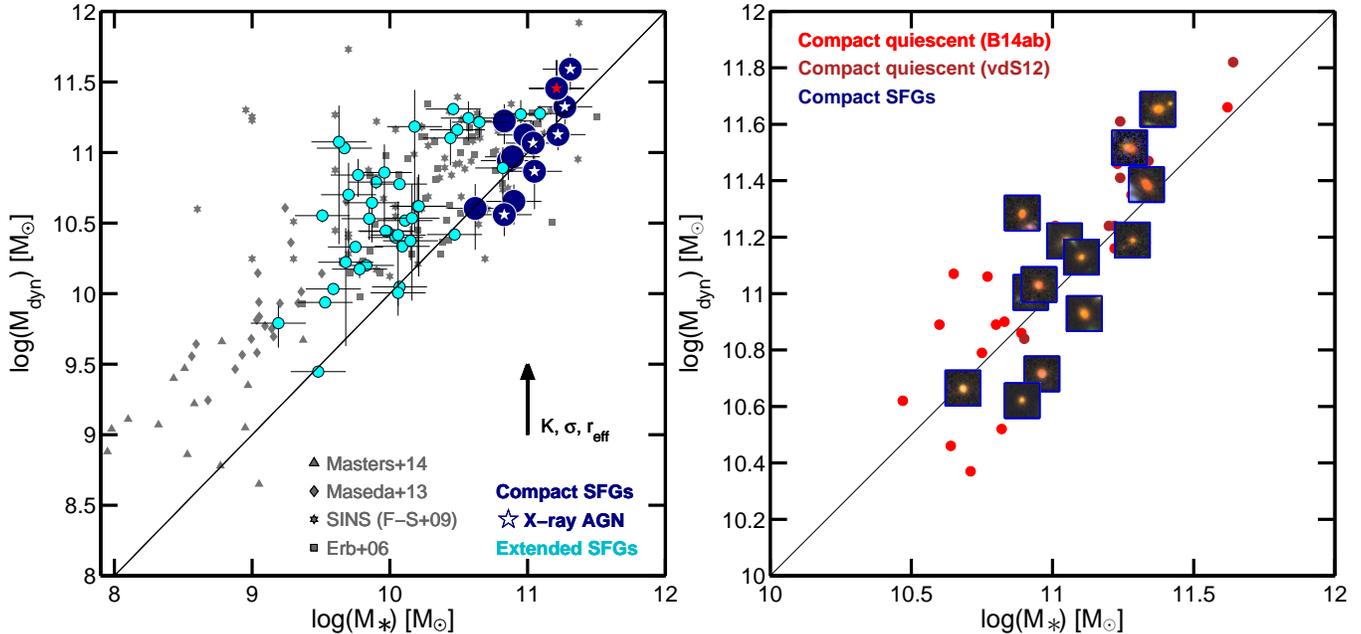}
\caption{\label{mdyn} {\it Left:} Stellar mass vs. dynamical mass the
  same galaxies shown in Figure~\ref{faberjackson}. The markers and
  colors indicate the same. The uncertainties add in quadrature a
  contribution from the statistical errors in size, stellar mass and
  \sigint~and the systematic uncertainty on the virial factor. We
  compare our measurements with the samples of intermediate and low
  mass SFGs in \cite{maseda13} and \cite{masters14}, and with the more
  massive galaxies in \citet{erb06} and the SINS survey
  \citep{fs09}. The arrow indicates the direction in which galaxies
  would move in the diagram as a function of a increasing virial
  constant $K$, \sigint~or $r_{\rm{e}}$. {\it Right:} Same as left
  panel but focusing on the high-mass end and showing only the $zJH$
  color postage stamps of compact SFGs and the distribution of compact
  quiescent galaxies drawn from \citet{vandesande13} and
  \citet{belli14a,belli14b}. Compact SFGs lie in the locus of compact
  quiescent galaxies suggesting that not only they have similar
  \mdyn$\sim$\mstar, but also consistent integrated kinematics
  (Figure~\ref{faberjackson}) and sizes
  (\mdyn$\sim$$\sigma_{\rm{int}}^{2}r_{\rm{e}}$).}
\end{figure*}

\subsection{Dynamical Masses}\label{dynmass}

In this section we make the operational assumption that compact SFGs
are dispersion dominated galaxies in order to calculate dynamical
masses. This is motivated by their elevated S\'ersic indices
($n\sim3.4$) and the similarities in their structural and kinematic
properties with compact quiescent galaxies ($r_{\rm{e}}\sim$~1.5~kpc;
axis-ratio$\sim$0.75) for which this is the usual assumption (although
see \citealt{vdw11a} and \citealt{chang13} for evidence that some
massive quiescent galaxies may be disks).

For dispersion-dominated galaxies, a simple virial argument, allows
one to relate the dynamical mass to its velocity dispersion and
effective radius, through a virial factor, $K$, using
\begin{equation} 
\label{mdyneq}
M_{\rm{dyn}}(<r_{\rm{e}}) = K\frac{\sigma_{\rm{int}}^{2}r_{\mathrm{e}}} {G}
\end{equation}
the value of $K$ depending on the mass density profile, the velocity
anisotropy, or the shape of the gravitational potential
\citep{courteau14}. \citet{cappellari06} calibrated this value from
surface brightness distributions and integral field kinematics of
local ellipticals, finding a S\'ersic dependent factor that ranges
from $K=3.6 - 2.5$ for low ($n=2$) and high ($n=5.5$) S\'ersic
galaxies, respectively. Similarly, \citet{binney08} show that a factor
of $K=3.35$ is appropriate for a variety of geometries and
mass-distributions. Here we adopt $K=$2.5, widely used in the analysis
of compact quiescent galaxies at $z\sim2$ (e.g., \citealt{newman12};
\citealt{vandesande13}, \citealt{belli14a}). We also multiply by 2 the
value obtained with Equation~\ref{mdyneq} (i.e., $2K=5$) to indicate
the {\it total} \mdyn~instead of within $r_{\rm{e}}$.

If we were to assume that compact SFGs are rotation-dominated
Equation~\ref{mdyneq} remains the same, but \sigint~is used to infer
$v_{\rm{rot}}$ making further assumptions on the geometry and
inclination of the galaxy. In general, the virial factor in that case
can vary from $K\sim2-10$ depending on the available structural
information. For spatially unresolved galaxies, the typical values
range from $K\sim3-6$ (e.g., \citealt{shapley04}; \citealt{erb06};
\citealt{maseda13}; \citealt{masters14}; also including the $\times$2
for the {\it total} \mdyn) bracketing our value within a factor of a
few. We use the same value for our extended SFGs to have a reference
sample for which \mdyn~is measured in the same way. To account for
the potential uncertainty in the assumed kinematic properties of
compact SFGs, we adopt a conservative error in the virial factor of
33\%.

\begin{figure*}[t]
\centering
\includegraphics[width=8.8cm,angle=0.]{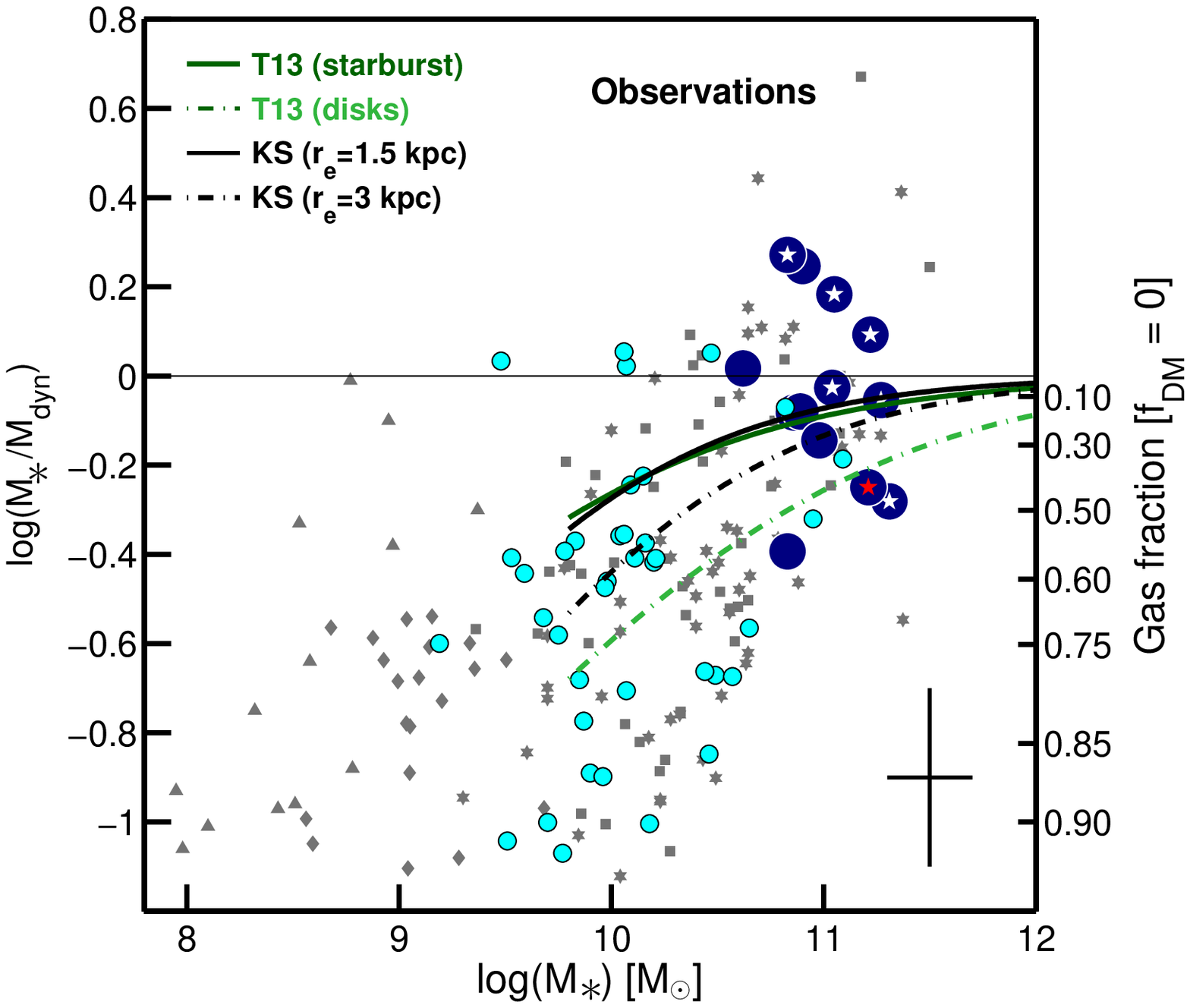}
\includegraphics[width=8.5cm,angle=0.]{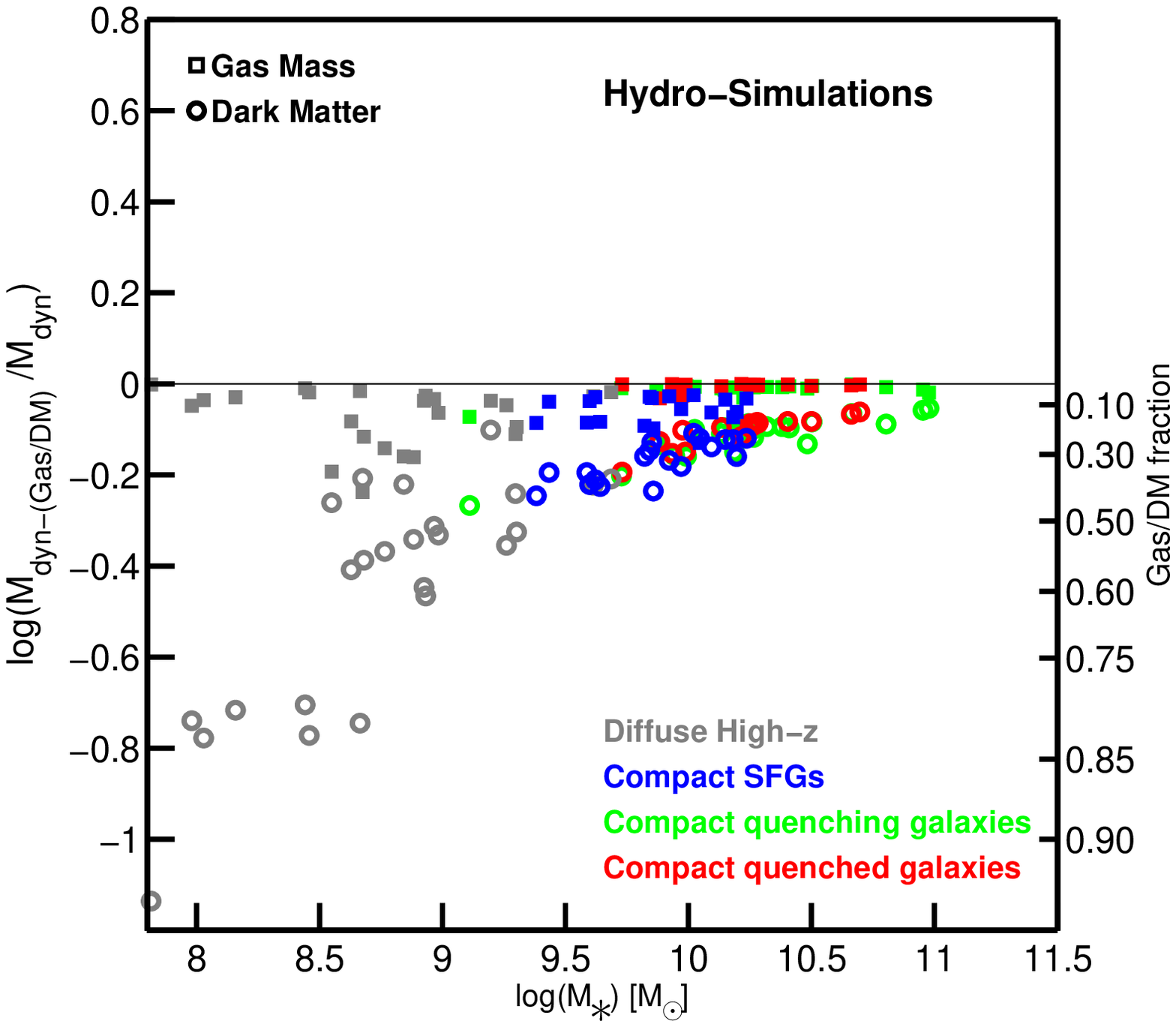}
\caption{\label{gasfraction} {\it Left:} Relative offset from the
  \mdyn$=$\mstar~relation (i.e., stellar mass fraction) as a function
  of stellar mass for compact and extended SFGs. The markers and the
  color scheme are the same as in Figure~\ref{mdyn}.  The right y-axis
  shows the inferred gas fraction assuming a negligible dark matter
  fraction, \mdyn$=$\mstar$+$\mgas. The error bars in the bottom-right
  corner indicate the average uncertainty in \mdyn~and \mstar~for
  compact and extended SFGs. The colored lines show the estimated gas
  fraction as a function of stellar mass from: (black) the KS relation
  \citep{ken98} assuming either a small, $r_{\rm{e}}=1.5$ (solid), or
  a large size, $r_{\rm{e}}=3$ (dashed-dotted); (green) the log-linear
  KS relation of \citet{tacconi13} for low, $\alpha_{\rm{CO}}=4.36$
  (light dashed-dotted), and high, $\alpha_{\rm{CO}}=1$ (dark solid),
  star-formation efficiencies typical of disk and starburst galaxies,
  respectively.  Compact SFGs may have small gas fractions of
  $\sim13\%$ due to their larger masses, smaller sizes and/or higher
  SFEs. {\it Right:} Estimated gas (square) and dark matter (circle)
  fractions in a sample of 30 hydro-dynamical simulations from
  \citet{ceverino10} and \citet{dekel13a}. The color code indicates
  different evolutionary stages in the simulations starting from
  diffuse low-mass galaxies at $z=4-5$ (gray) to compact SFGs (blue),
  compact quenching galaxies (green), and finally quiescent galaxies
  (red). The overall trend illustrates a similar decline in the gas
  and dark matter fractions with stellar mass (i.e., evolutionary
  stage) as in the observed galaxies. Interestingly, the models
  predict that the dark matter content dominates over the gas by a
  factor of $\sim3-5$ even in the most gas rich phases.}
\end{figure*}

Figure~\ref{mdyn} compares the stellar and dynamical mass estimates
for our sample of compact and extended SFGs, and other references from
the literature spanning a broad range of stellar masses from
\lmass$=8-11$. Despite the $\Delta$\lratio$\sim$0.3~dex scatter, the
evolution in \lratio~ appears to be consistent with a declining trend
as a function of stellar mass. This is clear in
Figure~\ref{gasfraction} which shows the relative offset from the 1:1
relation (i.e., the stellar mass fraction) ranging from
\lratio$=-0.57$~dex ($\sim$25\%) at lower masses to
\lratio$=-0.36$~dex ($\sim$45\%) in the most massive galaxies of
\citet{fs09}, and \lratio$=-0.06^{+0.21}_{-0.13}$~dex
(87$^{+13}_{-18}$\%) in compact SFGs. This suggest that \mstar~is the
main contributor to the dynamical mass in compact SFGs, and thus they
have only small gas or dark matter fractions. Moreover, the right
panel of Figure~\ref{mdyn} shows that, not only compact SFGs and
quiescent galaxies have similar integrated kinematics, but also
similar, stellar-dominated, \mdyn~, which provides indirect evidence
for the good agreement of their $r_{\rm{e}}$
(\mdyn$\sim$$\sigma_{\rm{int}}^{2}r_{\rm{e}}$).

Note that while the trend in Figure~\ref{mdyn} depends on the virial
factor, our choice of $K=5$ is typically larger than the usual value
adopted for extended SFGs (except for the few low-mass
dispersion-dominated galaxies in \citealt{fs09}; $K=6.7$). Therefore,
lowering $K$ for compact SFGs, or increasing it for extended SFGs,
would only increase the relative difference, suggesting that the mass
dependence on \lratio~ is not an artifact of the virial factor.

\subsection{Gas and dark matter fractions in compact SFGs}

Assuming that $\sigma_{\rm{int}}$ is dynamical in origin, the offset
between \mdyn~and \mstar~can be interpreted as evidence for other
components contributing to the gravitational potential of the galaxy,
such as dark matter or gas (molecular and atomic):
M$_{\rm{dyn}}=M_{\star}+M_{\mathrm{gas}}+M_{\rm{DM}}$.  In the absence
of more direct measurements of the molecular gas content from CO or
far-IR estimates (e.g., \citealt{daddi10a}; \citealt{magdis12}) it is
unclear what the breakout is between the two contributions. In
\S~\ref{gasfrac} we estimate the gas fraction if we assume that the
offset in \mdyn~is mainly driven by the gas content. This is a
plausible assumption given that galaxies at $z\sim2$ are known to have
large gas reservoirs (\citealt{tacconi10,tacconi13}), and they could
still be accreting gas from their parent halos (\citealt{keres05};
\citealt{dekel06}; \citealt{dekel09a}). Furthermore, if massive
compact galaxies become the core of local ellipticals
\citep{bezanson09}, their dark matter fractions should be relatively
small ($\sim$10\%; e.g., \citealt{cappellari13}). Nevertheless, given
the assumptions above, the estimated gas fractions should be regarded
as upper limits. In section~\ref{dmfrac} we also comment on the
contribution of dark matter to \mdyn~from the predictions of
theoretical simulations.

\subsubsection{Inferred gas fractions}\label{gasfrac}

If the offset in \mdyn~vs. \mstar~reflects primarily gas content,
Figure~\ref{gasfraction} suggests that the gas fraction in SFGs
declines with stellar mass from \lgas$\sim0.73$ in low-mass galaxies
to \lgas$=0.13^{+0.17}_{-0.13}$ in compact SFGs. This trend is roughly
consistent with that of \citet{tacconi13}, based on CO observations,
who find a decrease in the gas fraction from \lgas$=0.60$ to $0.35$ in
the stellar mass range \lmass$=10.4-11.2$. However, at the high-mass
end, our results are a factor of $\sim$2 smaller than the CO estimates
(dashed-dotted green line in Figure~\ref{gasfraction}), as also noted
by \citet{fs09} for the SINS sample. This difference can be partially
due to the large scatter and the systematic uncertainties in the
virial factor or the $\alpha_{\rm{CO}}$ conversion (e.g.,
\citealt{genzel10}). However, we speculate that it can be caused by
differences in the structural properties or star-formation
efficiencies of the galaxies.

Indeed, the estimated gas fractions from the Kennicutt-Schmidt
relation (KS) between the gas and star-formation rate densities
\citep{ken98}:
\begin{equation}\label{ksr}
\rm{log}(M_{\rm{gas}})= 8.715 + 0.71 \rm{log}(\rm{SFR}) + 0.57 log(r_{\rm{e}})
\end{equation}
implies that compact SFGs on the main-sequence, i.e, those having
similar SFRs and stellar masses as other (extended) SFGs
(Figure~\ref{selectiondiag}, also Barro+14), have lower M$_{\rm{gas}}$
simply because their sizes are up to 5 times smaller.  Using the best
fit to the massive (\lmass$>10$) main-sequence shown in
Figure~\ref{selectiondiag} ($0.41($\lmass$-10.5)+1.8$; see also
\citealt{whitaker13} for a similar result) we estimate the gas
fraction as a function of stellar mass and size from
Equation~\ref{ksr}. The black lines in Figure~\ref{gasfraction} show
evolution of \lgas for the average size of extended
($r_{\rm{e}}\sim3$~kpc) and compact SFGs ($r_{\rm{e}}\sim1.5$~kpc),
suggesting that the latter have $\sim$15\% lower gas fractions than
the former. In fact, the prediction for compact SFGs, \lgas$=0.10$, is
roughly consistent with the estimate from the kinematic measurements.

Similarly, we can estimate the gas fraction following the empirical KS
relation from \citet{tacconi13}. In this case, the authors assume a
log-linear relation, which factors out the size dependence, to provide
an estimate of the star-formation efficiency (SFE) or gas depletion
timescale,
$\Sigma_{\rm{SFR}}/\Sigma_{\rm{gas}}=$SFR/M$_{\rm{gas}}$$=$1/t$_{\rm{dep}}\equiv$SFE.
Note however that the empirical relation depends on the assumed value
of $\alpha_{\rm{CO}}$, which relates the observed L$_{\rm{CO}}$ to
M$_{\rm{gas}}$. In \citet{tacconi13} the authors use the Milky Way
conversion factor ($\alpha_{\rm{CO}}=4.36$) which is appropriate for a
broad range of {\it normal} SFGs between $z\sim1-2$
(\citealt{daddi08}; \citealt{daddi10a}; \citealt{genzel12}). But this
factor may be smaller in galaxies with higher SFEs similar to local
(U)LIRGs or sub-mm galaxies ($\alpha_{\rm{CO}}\lesssim1$;
\citealt{Solomon05}; \citealt{tacconi08}; \citealt{engel10}). Several
papers have indeed pointed out the presence of a population of more
efficient ``starburst'' galaxies which are typically identified by
elevated SFRs above the average main sequence (\citealt{daddi10b};
\citealt{rodi11}; \citealt{genzel10}; \citealt{magnelli12a}). In such
galaxies, the gas depletion timescales may change drastically,
decreasing from $t_{\rm{dep}}=0.7$~Gyr in star-forming disks
(\citealt{tacconi13}; \citealt{magdis12}) to
$t_{\rm{dep}}=0.2-0.3$~Gyr.

Interestingly, the fraction of high efficiency ``starburst'' among
massive SFGs seems to depend on the galaxy structure, such that the
fraction increases among compact (or high \sigint) galaxies
(\citealt{daddi10a}; \citealt{elbaz11}; \citealt{narayanan12}). If
that is the case, compact SFGs have lower gas fractions than other
(extended) SFGs due to a higher SFE. The green lines in
Figure~\ref{gasfraction} show the evolution of the gas fraction as a
function of \mstar~following the KS relation of \citet{tacconi13} for
low (disk-like) and a high (``starburst'') SFEs. In this case, the
difference between the two possibilities is slightly larger. Within
the large scatter, the relation for disks seems to describe well the
observed distribution for extended SFGs, and the samples from the
references, while the ``starburst'' relation seems to better match the
estimated gas fractions for compact SFGs.

\subsubsection{Comparison to Hydrodynamical simulations}\label{dmfrac}

In absence of 2D-kinematic data and gas measurements to gauge the
expected contribution of dark matter and molecular gas to \mdyn,
theoretical models can provide some insight on the relative
contribution of each of these components. The right panel of
Figure~\ref{gasfraction} shows the dark matter (circles) and gas
(squares) fractions as a function of the stellar mass for a small
sample of simulated galaxies at different stages of their
evolution. These galaxies are drawn from the larger sample of
\citet{ceverino10,ceverino14} and \citet[][also Zolotov et al. in
  prep]{dekel13a}, computed with the Adaptive Refinement Tree (ART)
code (\citealt{arm}) using a spatial resolution of $\sim$25~pc (see
\citet{ceverino09} for more details about the code).

As shown in \citet{barro14} and \citet{dekel13b} these simulated
galaxies appear to describe well the formation of compact SFGs from
more extended star-forming precursors that experience a wet
contractions as a result of mergers, violent disk instabilities or a
combination of both. The color-code on the Figure illustrates the
evolutionary sequence from high-redshift low-mass galaxy seeds to
extended star-forming galaxies on the verge of (or experiencing) a
dissipational contraction that will transform them into compact SFGs,
and finally quenching(-ed) galaxies running out of gas.  The blue
markers indicate approximately the simulated galaxies at the peak of
the compact phase. However, since the time spent at this phase is
short, some of them may already be on the (green) quenching
group. Note also that simulated galaxies are, by selection, somewhat
less massive than the observed galaxies at the compact phase
\citep{dekel13a}.

The overall trend in the simulations is qualitatively similar to the
observations, illustrating the formation of a dense stellar core that
becomes self-gravitating (i.e., \mstar~$>$~\mdm) at the expense of
turning gas into stars in the innermost region of
galaxies. Interestingly, models predict the dominance of dark matter
over gas even at the low mass end where galaxies are more gas rich
(15\% vs. 50\%).  In extended galaxies turning into compact, a
dissipational wet inflow causes gas (and stars) to migrate inward
reaching a steady state in which gas does not accumulate in the center
but slowly decreases being turned into stars (``bathtub''), thus
diluting the dark matter fraction until it reaches a base value of
$\sim$10\% (see also Zolotov et al. 2014 in prep.).  Although the exact
gas and dark matter fractions vary from galaxy to galaxy the overall
trend suggests that the dark matter content in compact SFGs is not
negligible, implying that the gas fractions inferred in the previous
section could be even smaller.

\subsubsection{Quenching timescales for compact SFGs}

The close agreement among \mdyn~and \mstar~in compact SFGs implies
that their gas and dark matter content must be small. The large
uncertainties in \mdyn~prevent a better constraint on the gas
fraction, but as discussed above, unless all compact SFGs have rare
intrinsic kinematics (e.g., if they are all edge-on disks), \mdyn~is
not underestimated by more than a factor of $\times1.3-1.5$ (i.e.,
$K=7-8$). Furthermore, most of the other complications such as an
underestimated value of $r_{\rm{e}}$, or an overestimated
\sigint~would move the galaxies towards smaller \mdyn~and therefore
gas fractions.

Taking the 13\% gas fraction calculated in \S~\ref{gasfrac} at face
value, and assuming no further gas accretion onto the galaxy, compact
SFGs would consume their current reservoirs on timescales of
t$_{\rm{dep}}$=M$_{\rm{gas}}$/SFR$=$230$^{+110}_{-190}$~Myr, where the
error bars reflect the uncertainty in \mgas~ and SFR, as well as the
small number statistics. This value fits well in the range of
predicted quenching timescales required for compact SFGs to reproduce
the increase in the number density of compact quiescent galaxies from
$z=3$ to $z=2$ (\citealt{barro13}). Note also, that a larger value of
$K$ does not necessarily imply longer quenching times if, as predicted
by the simulations, the dark matter fraction is larger than the gas
fraction in the inner regions of compact SFGs.

Most of the uncertainties in \mdyn~arise from the poor constrains on
the kinematic properties of the galaxies. This problem shows the
obvious need for a detailed analysis of their {\it resolved}
kinematics to quantify whether they are rotating disks or dispersion
dominated galaxies. Surveys of Integral Field Spectroscopy aided by
adaptive optics (\citealt{law09}; \citealt{snewman13}; Forster
Schreiber et al. 2014 in prep.) or high spatial resolution sub-mm
observations with ALMA (e.g., \citealt{gilli14}; \citealt{debreuck14})
will shed some light on the resolved dynamics of compact SFGs
providing more precise estimates of their \mdyn~and gas fractions.

\subsection{Outflows from Extended Gas Regions?}\label{outflows}

The unusual formation processes responsible for compact star-forming
galaxies might leave an imprint in their extended gas halos.  For
example, violent assembly by mergers typically leads to large
($>$10~kpc) gas halos (\citealt{hernquist95}; \citealt{dokkum10b}).
Similarly, feedback from the X-ray AGNs frequently present in compact
star-forming galaxies (\citealt{barro13}; \citealt{rangel14}) would
produce large extended halos of partially ionized gas, as observed
around nearby quasars \citep{fu09}.  While we have not found {\it
  kinematic} evidence for inflowing or outflowing gas, the gas halos
might be kinematically relaxed but still visible as spatially extended
emission lines.

\begin{figure}[t]
\includegraphics[width=9cm,angle=0.]{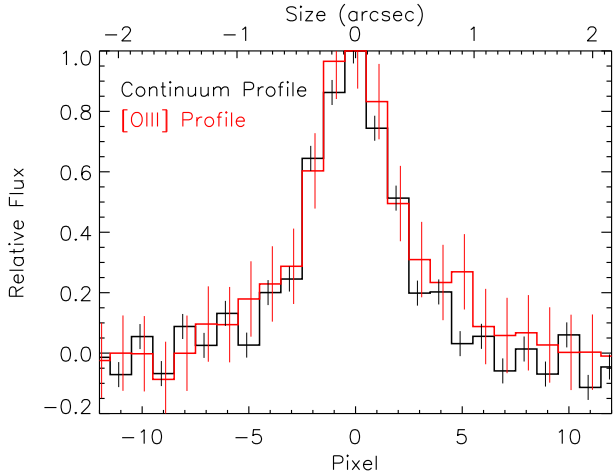}
\caption{\label{extended} Spatial profiles of stacked spectra for both
  the continuum and $\OIII\lambda$5007 emission line.  In both cases,
  the galaxies are marginally resolved, with
  FWHM$\sim$5~pix$\sim$0$\farcs$9: larger than the typical seeing of
  $0\farcs5 - 0\farcs7$.  The $\OIII$ emission line has very slightly
  more flux in the wings of the spatial profile, although the widths
  of each profile are not significantly different.  In other words,
  our data do not show any obvious signs of a massive gas reservoir at
  large radii deposited by feedback or inflows.}
\end{figure}

We search for evidence of extended emission line regions in the
compact SF galaxies by comparing the spatial extent of the
$\OIII\lambda$5007 emission line with the continuum.  For sufficient
signal-to-noise, we stack the spectra of compact SFGs with
well-detected $\OIII$ lines (IDs 20659, 23896, 25998, 23382, 25952,
26211, 9218).  The stacked two-dimensional spectrum is constructed by
weighting each pixel in each object by the inverse of its error.  The
$\OIII$ region is defined by $5004<\lambda_{\rm{rest}}<5010$, and the
continuum by $4750<\lambda_{\rm{rest}}<4851$,
$4871<\lambda_{\rm{rest}}<4949$, $4969<\lambda_{\rm{rest}}<4997$, and
$5017<\lambda_{\rm{rest}}<5120$.  The spatial profiles of the $\OIII$
emission line and continuum are shown in Figure~\ref{extended}.

The full-width at half-maxima (FWHMs) of the two profiles are
statistically consistent with one another: $5.5 \pm 0.5$~pixels for
$\OIII$ and $5.2 \pm 0.2$~pixels for the continuum.  Both translate to
about $0\farcs9$ (using the MOSFIRE pixel scale of
$0\farcs18$~pix$^{-1}$), slightly larger than the typical seeing
($0\farcs5-0\farcs7$) of the observations, which indicates that (at
least some of) the galaxies are resolved in the MOSFIRE spectra.  The
first moment of the $\OIII$ line ($2.24\pm0.21$~pixels) is marginally
larger than the continuum ($1.94\pm0.06$~pixels), influenced by the
excess $\OIII$ flux in the wings of the profile (at $0\farcs5-1''$, or
$\sim$5-10~kpc at $z \sim 2$).  However the line width difference is
not statistically significant, and thus we conclude that our compact
star-forming galaxies show no obvious evidence of spatially extended
excited gas.

\section{Summary}

Using the Keck-I MOSFIRE infrared spectrograph we measure integrated
velocity dispersions from the emission lines of 13 massive
(\lmass$\sim$10.8), dusty (IR-bright), compact SFGs
($\Sigma_{1.5}>10.4$) at redshift $2\leq z \leq2.5$ to investigate an
evolutionary connection between them and compact quiescent galaxies at
$z\sim2$.

Compact SFGs have large velocity dispersions of
\sigint$=230^{+40}_{-30}$~km s$^{-1}$ consistent with the
absorption-line-based measurements for a sample of equally massive
quiescent galaxies at $z\sim2$, and both populations follow a similar,
tight ($\Delta$log(\sigint)$=$0.14~dex), \mstar$-$\sigint~relation
over $\sim$1~dex in stellar mass. For one compact SFG in common with
\citet{belli14b}, the gas and stellar velocity dispersions are
consistent at a 1$\sigma$ level suggesting that the width of the
emission lines traces the gravitational potential. The dynamical
masses of compact SFGs and quiescent galaxies, are also in excellent
agreement and present a small offset with respect to the
\mdyn$=$\mstar~relation, \lratio$=-0.06\pm0.2$~dex, which contrast
with the larger deviation found in other (extended) SFGs at the same
redshift. These results suggest that: 1) compact SFGs are
kinematically relaxed, i.e., the dispersion is gravitational in
origin; 2) the stellar component dominates the gravitational
potential, and thus compact SFGs have only small gas or dark matter
fractions. In the absence of dark matter and further gas accretion,
the average gas fraction in compact SFGs ($13^{+17}_{-13}$\%) imply
short depletion timescales, t$_{\rm{dep}}=230^{+110}_{-190}$~Myr.

The excellent agreement in the stellar and dynamical masses,
structural properties and kinematic properties of compact SFGs and
quiescent galaxies, jointly with the a priori short quenching
timescales, provide further support to the evolutionary sequence
proposed in \citet[][also Dekel\&Burkert 2014]{barro13,barro14}, in
which compact SFGs are the immediate progenitors of compact quiescent
galaxies at $z\sim2$.

This is the first observational effort aimed at studying the
kinematics of massive {\it compact} SFGs, which represent a key phase
in the formation of quiescent galaxies, but are typically absent from
current spectroscopic surveys at $z\sim2$. Our results open the door
to a better characterization of these galaxies, which would require
direct measurements of their intrinsic kinematics, currently
unresolved in seeing-limited observations. This can be achieved either
with AO-assisted NIR spectroscopy or high-resolution sub-mm
observations with ALMA.

\section*{Acknowledgments}

We thank Sirio Belli for useful discussions and Dan Masters for
providing us with additional data for his galaxies. Support for
Program number HST-GO-12060 was provided by NASA through a grant from
the Space Telescope Science Institute, which is operated by the
Association of Universities for Research in Astronomy, Incorporated,
under NASA contract NAS5-26555. GB acknowledges support from NSF grant
AST-08-08133. PGP-G acknowledges support from grant AYA2012-31277-E.
This work has made use of the Rainbow Cosmological Surveys Database,
which is operated by the Universidad Complutense de Madrid (UCM),
partnered with the University of California Observatories at Santa
Cruz (UCO/Lick,UCSC). CP is supported by the KASI-Yonsei Joint
Research Program (2014) for the Frontiers of Astronomy and Space
Science funded by the Korea Astronomy and Space Science Institute. The
authors recognize and acknowledge the very significant cultural role
and reverence that the summit of Mauna Kea has always had within the
indigenous Hawaiian community. We are most fortunate to have the
opportunity to conduct observations from this mountain.

%Table with properties.
%\placetable{redshifts}
%\clearpage
%\LongTables
%\begin{landscape}
\begin{deluxetable*}{cccccccccccccc}
%\tablecolumns{8}
%\rotate
%\rotate{90}
\setlength{\tabcolsep}{0.002in} 
\tablewidth{0pt}
\tabletypesize{\scriptsize}
\tablecaption{\label{redshifts} Stellar and spectroscopic properties of compact SFGs}
\tablehead{
\colhead{ID}  & \colhead{R.A.} & \colhead{DEC} & \colhead{$z_{\mathrm{spec}}$} & \colhead{log~M$_{\star}$} &
 \colhead{$f_{24\mu m}$} & \colhead{$f_{100\mu m}$}  & \colhead{$f_{160\mu m}$}  &
\colhead{SFR}  &\colhead{r$_{eff}$} &  \colhead{$\sigma_{\rm{int}}$} &
\colhead{\NII/\Ha}   &  \colhead{log~M$_{\rm{dyn}}$} &  \colhead{$L_{\rm{X}}$}\\
\colhead{(1)} & \colhead{(2)}  & \colhead{(3)} & \colhead{(4)}   & \colhead{(5)}   & 
\colhead{(6)}  & \colhead{(7)}  & \colhead{(8)}   & \colhead{(9)}  &
\colhead{(10)} & \colhead{(11)} & \colhead{(12)} &
\colhead{(13)} & \colhead{(14)}}
\startdata
20659& 53.182839& -27.734911& 2.432& 10.90& 72& -& -& 96& 1.00$\pm$0.01& 197$\pm$37& 0.77$\pm$0.30& 10.65$\pm$0.26& -\\
23896& 53.100814& -27.715986& 2.303& 10.86& 49& -& -& 57& 1.76$\pm$0.05& 207$\pm$33& 0.17$\pm$0.59& 10.94$\pm$0.23& 0.24\\
25998& 53.137572& -27.700104& 2.453& 10.89& 140& 3359& 6716& 158& 1.19$\pm$0.03& 260$\pm$18& 1.70$\pm$0.78& 10.97$\pm$0.10& -\\
23382& 53.162299& -27.712135& 2.433& 11.27& 80& -& 2175& 102& 2.01$\pm$0.03& 300$\pm$57& -& 11.32$\pm$0.27& 0.20\\
25952& 53.121136& -27.698075& 1.970& 10.62& 91& -& -& 48& 0.94$\pm$0.02& 192$\pm$55& 0.57$\pm$0.49& 10.60$\pm$0.41& -\\
26211& 53.065952& -27.701852& 2.154& 10.83& 153& 1738& -& 103& 1.40$\pm$0.02& 320$\pm$59& -& 11.22$\pm$0.26& -\\
8846& 189.026369& 62.209125& 2.487& 11.22& 115& 1477& 6057& 135& 1.95$\pm$0.14& 243$\pm$30& 0.73$\pm$0.18& 11.13$\pm$0.19& 1.68\\
4374& 189.028576& 62.172614& 2.321& 11.31& 107& -& 6858& 139& 2.04$\pm$0.03& 406$\pm$69& 2.17$\pm$0.28& 11.59$\pm$0.24& 0.21\\
12605& 189.087068& 62.237622& 2.090& 10.98& 96& 832& -& 53& 2.91$\pm$0.10& 198$\pm$31& 0.74$\pm$0.18& 11.12$\pm$0.22& -\\
17409& 189.182966& 62.272470& 2.322& 11.05& 55& 1181& 2833& 316& 3.19$\pm$0.02& 141$\pm$38& 0.89$\pm$0.39& 10.87$\pm$0.38& 0.51\\
14548& 189.251901& 62.252460& 2.330& 11.04& 84& -& 1937& 79& 2.02$\pm$0.06& 223$\pm$56& 0.92$\pm$0.30& 11.07$\pm$0.35& 0.74\\
9218& 189.260847& 62.212224& 2.420& 10.83& 64& -& -& 54& 1.28$\pm$0.03& 156$\pm$27& 0.23$\pm$0.39& 10.56$\pm$0.24& 2.47\\
10289& 150.074608 & 2.302008& 2.095& 11.21& 220& -& -& 110& 2.09$\pm$0.05& 352$\pm$213& $<0.5$& 11.39$\pm$045& 1.67
\enddata                                                                    
\tablecomments{\\ 
(1) General ID in the CANDELS $H$-band selected catalog in GOODS-S (Guo et al. 2013), GOODS-N (Barro et al. in prep.) and COSMOS (Nayyeri et al. in prep.) catalogs .\\
(2,3) R.A and Declination J2000.\\
(4) Spectroscopic redshift.\\
(5) Stellar mass (\lmass) determined from SED fitting using \citet{bc03}.
(6,7,8) Far-IR fluxes in {\it Spitzer}/MIPS~24~$\mu$m and {\it Herschel}/PACS~100~$\mu$m and PACS~160~$\mu$m.\\
(9) Total star formation rate (SFR$_{UV+IR}$ [M$_{\odot}$yr$^{-1}$]), see \S~\ref{stellarprop}.\\
(10) Circularized, effective (half-light) radius (kpc) measured with GALFIT, see \S~\ref{stellarprop}.\\
(11) Integrated velocity dispersion measure from the line width (FWHM), see \S~\ref{kinmes}.\\
(12) \NII~ to \Ha~ line ratio.\\
(13) Dynamical mass estimated from the velocity dispersion and effective radius of the galaxy, see \S~\ref{dynmass}.\\
(14) Full band X-ray luminosity in units of 10$^{44}$ erg s$^{-1}$.\\
\\}                                                                         
\end{deluxetable*}

\bibliographystyle{aa}
\bibliography{referencias}

\begin{thebibliography}{129}
\expandafter\ifx\csname natexlab\endcsname\relax\def\natexlab#1{#1}\fi

\bibitem[{{Alexander} {et~al.}(2003){Alexander}, {Bauer}, {Brandt},
  {Schneider}, {Hornschemeier}, {Vignali}, {Barger}, {Broos}, {Cowie},
  {Garmire}, {Townsley}, {Bautz}, {Chartas}, \& {Sargent}}]{chandra2m}
{Alexander}, D.~M., {Bauer}, F.~E., {Brandt}, W.~N., {et~al.} 2003, \aj, 126,
  539

\bibitem[{{Baldwin} {et~al.}(1981){Baldwin}, {Phillips}, \& {Terlevich}}]{bpt}
{Baldwin}, J.~A., {Phillips}, M.~M., \& {Terlevich}, R. 1981, \pasp, 93, 5

\bibitem[{{Barro} {et~al.}(2013){Barro}, {Faber}, {P{\'e}rez-Gonz{\'a}lez},
  {Koo}, {Williams}, {Kocevski}, {Trump}, {Mozena}, {McGrath}, {van der Wel},
  {Wuyts}, {Bell}, {Croton}, {Ceverino}, {Dekel}, {Ashby}, {Cheung},
  {Ferguson}, {Fontana}, {Fang}, {Giavalisco}, {Grogin}, {Guo}, {Hathi},
  {Hopkins}, {Huang}, {Koekemoer}, {Kartaltepe}, {Lee}, {Newman}, {Porter},
  {Primack}, {Ryan}, {Rosario}, {Somerville}, {Salvato}, \& {Hsu}}]{barro13}
{Barro}, G., {Faber}, S.~M., {P{\'e}rez-Gonz{\'a}lez}, P.~G., {et~al.} 2013,
  \apj, 765, 104

\bibitem[{{Barro} {et~al.}(2014){Barro}, {Faber}, {Perez-Gonzalez}, {Pacifici},
  {Trump}, {Koo}, {Wuyts}, {Guo}, {Bell}, {Dekel}, {Porter}, {Primack},
  {Ferguson}, {Ashby}, {Caputi}, {Ceverino}, {Croton}, {Fazio}, {Giavalisco},
  {Hsu}, {Kocevski}, {Koekemoer}, {Kurczynski}, {Kollipara}, {Lee}, {McIntosh},
  {McGrath}, {Moody}, {Somerville}, {Papovich}, {Salvato}, {Santini},
  {Williams}, {Willner}, \& {Zolotov}}]{barro14}
{Barro}, G., {Faber}, S.~M., {Perez-Gonzalez}, P.~G., {et~al.} 2014, ArXiv
  e-prints

\bibitem[{{Bell} {et~al.}(2005){Bell}, {Papovich}, {Wolf}, {Le Floc'h},
  {Caldwell}, {Barden}, {Egami}, {McIntosh}, {Meisenheimer},
  {P{\'e}rez-Gonz{\'a}lez}, {Rieke}, {Rieke}, {Rigby}, \& {Rix}}]{bell05}
{Bell}, E.~F., {Papovich}, C., {Wolf}, C., {et~al.} 2005, \apj, 625, 23

\bibitem[{{Bell} {et~al.}(2012){Bell}, {van der Wel}, {Papovich}, {Kocevski},
  {Lotz}, {McIntosh}, {Kartaltepe}, {Faber}, {Ferguson}, {Koekemoer}, {Grogin},
  {Wuyts}, {Cheung}, {Conselice}, {Dekel}, {Dunlop}, {Giavalisco},
  {Herrington}, {Koo}, {McGrath}, {de Mello}, {Rix}, {Robaina}, \&
  {Williams}}]{bell12}
{Bell}, E.~F., {van der Wel}, A., {Papovich}, C., {et~al.} 2012, \apj, 753, 167

\bibitem[{{Belli} {et~al.}(2014{\natexlab{a}}){Belli}, {Newman}, \&
  {Ellis}}]{belli14a}
{Belli}, S., {Newman}, A.~B., \& {Ellis}, R.~S. 2014{\natexlab{a}}, \apj, 783,
  117

\bibitem[{{Belli} {et~al.}(2014{\natexlab{b}}){Belli}, {Newman}, {Ellis}, \&
  {Konidaris}}]{belli14b}
{Belli}, S., {Newman}, A.~B., {Ellis}, R.~S., \& {Konidaris}, N.~P.
  2014{\natexlab{b}}, ArXiv e-prints

\bibitem[{{Bennert} {et~al.}(2002){Bennert}, {Falcke}, {Schulz}, {Wilson}, \&
  {Wills}}]{bennert02}
{Bennert}, N., {Falcke}, H., {Schulz}, H., {Wilson}, A.~S., \& {Wills}, B.~J.
  2002, \apjl, 574, L105

\bibitem[{{Bezanson} {et~al.}(2013){Bezanson}, {van Dokkum}, {van de Sande},
  {Franx}, \& {Kriek}}]{bezanson13}
{Bezanson}, R., {van Dokkum}, P., {van de Sande}, J., {Franx}, M., \& {Kriek},
  M. 2013, \apjl, 764, L8

\bibitem[{{Bezanson} {et~al.}(2009){Bezanson}, {van Dokkum}, {Tal},
  {Marchesini}, {Kriek}, {Franx}, \& {Coppi}}]{bezanson09}
{Bezanson}, R., {van Dokkum}, P.~G., {Tal}, T., {et~al.} 2009, \apj, 697, 1290

\bibitem[{{Binney} \& {Tremaine}(2008)}]{binney08}
{Binney}, J. \& {Tremaine}, S. 2008, {Galactic Dynamics: Second Edition}
  (Princeton University Press)

\bibitem[{{Bothwell} {et~al.}(2013){Bothwell}, {Smail}, {Chapman}, {Genzel},
  {Ivison}, {Tacconi}, {Alaghband-Zadeh}, {Bertoldi}, {Blain}, {Casey}, {Cox},
  {Greve}, {Lutz}, {Neri}, {Omont}, \& {Swinbank}}]{bothwell13}
{Bothwell}, M.~S., {Smail}, I., {Chapman}, S.~C., {et~al.} 2013, \mnras, 429,
  3047

\bibitem[{{Brammer} {et~al.}(2008){Brammer}, {van Dokkum}, \& {Coppi}}]{eazy}
{Brammer}, G.~B., {van Dokkum}, P.~G., \& {Coppi}, P. 2008, \apj, 686, 1503

\bibitem[{{Brammer} {et~al.}(2011){Brammer}, {Whitaker}, {van Dokkum},
  {Marchesini}, {Franx}, {Kriek}, {Labb{\'e}}, {Lee}, {Muzzin}, {Quadri},
  {Rudnick}, \& {Williams}}]{brammer11}
{Brammer}, G.~B., {Whitaker}, K.~E., {van Dokkum}, P.~G., {et~al.} 2011, \apj,
  739, 24

\bibitem[{{Bruzual} \& {Charlot}(2003)}]{bc03}
{Bruzual}, G. \& {Charlot}, S. 2003, \mnras, 344, 1000

\bibitem[{{Buitrago} {et~al.}(2014){Buitrago}, {Conselice}, {Epinat},
  {Bedregal}, {Gr{\"u}tzbauch}, \& {Weiner}}]{buitrago14}
{Buitrago}, F., {Conselice}, C.~J., {Epinat}, B., {et~al.} 2014, \mnras, 439,
  1494

\bibitem[{{Buitrago} {et~al.}(2008){Buitrago}, {Trujillo}, {Conselice},
  {Bouwens}, {Dickinson}, \& {Yan}}]{buitrago08}
{Buitrago}, F., {Trujillo}, I., {Conselice}, C.~J., {et~al.} 2008, ArXiv
  e-prints

\bibitem[{{Buitrago} {et~al.}(2013){Buitrago}, {Trujillo}, {Conselice}, \&
  {H{\"a}u{\ss}ler}}]{buitrago13}
{Buitrago}, F., {Trujillo}, I., {Conselice}, C.~J., \& {H{\"a}u{\ss}ler}, B.
  2013, \mnras, 428, 1460

\bibitem[{{Calzetti} {et~al.}(2000){Calzetti}, {Armus}, {Bohlin}, {Kinney},
  {Koornneef}, \& {Storchi-Bergmann}}]{calzetti}
{Calzetti}, D., {Armus}, L., {Bohlin}, R.~C., {et~al.} 2000, \apj, 533, 682

\bibitem[{{Cappellari} {et~al.}(2006){Cappellari}, {Bacon}, {Bureau}, {Damen},
  {Davies}, {de Zeeuw}, {Emsellem}, {Falc{\'o}n-Barroso}, {Krajnovi{\'c}},
  {Kuntschner}, {McDermid}, {Peletier}, {Sarzi}, {van den Bosch}, \& {van de
  Ven}}]{cappellari06}
{Cappellari}, M., {Bacon}, R., {Bureau}, M., {et~al.} 2006, \mnras, 366, 1126

\bibitem[{{Cappellari} {et~al.}(2013){Cappellari}, {McDermid}, {Alatalo},
  {Blitz}, {Bois}, {Bournaud}, {Bureau}, {Crocker}, {Davies}, {Davis}, {de
  Zeeuw}, {Duc}, {Emsellem}, {Khochfar}, {Krajnovi{\'c}}, {Kuntschner},
  {Morganti}, {Naab}, {Oosterloo}, {Sarzi}, {Scott}, {Serra}, {Weijmans}, \&
  {Young}}]{cappellari13}
{Cappellari}, M., {McDermid}, R.~M., {Alatalo}, K., {et~al.} 2013, \mnras, 432,
  1862

\bibitem[{{Cassata} {et~al.}(2011){Cassata}, {Giavalisco}, {Guo}, {Renzini},
  {Ferguson}, {Koekemoer}, {Salimbeni}, {Scarlata}, {Grogin}, {Conselice},
  {Dahlen}, {Lotz}, {Dickinson}, \& {Lin}}]{cassata11}
{Cassata}, P., {Giavalisco}, M., {Guo}, Y., {et~al.} 2011, \apj, 743, 96

\bibitem[{{Cassata} {et~al.}(2013){Cassata}, {Giavalisco}, {Williams}, {Guo},
  {Lee}, {Renzini}, {Ferguson}, {Faber}, {Barro}, {McIntosh}, {Lu}, {Bell},
  {Koo}, {Papovich}, {Ryan}, {Conselice}, {Grogin}, {Koekemoer}, \&
  {Hathi}}]{cassata13}
{Cassata}, P., {Giavalisco}, M., {Williams}, C.~C., {et~al.} 2013, \apj, 775,
  106

\bibitem[{{Ceverino} {et~al.}(2010){Ceverino}, {Dekel}, \&
  {Bournaud}}]{ceverino10}
{Ceverino}, D., {Dekel}, A., \& {Bournaud}, F. 2010, \mnras, 404, 2151

\bibitem[{{Ceverino} \& {Klypin}(2009)}]{ceverino09}
{Ceverino}, D. \& {Klypin}, A. 2009, \apj, 695, 292

\bibitem[{{Ceverino} {et~al.}(2014){Ceverino}, {Klypin}, {Klimek},
  {Trujillo-Gomez}, {Churchill}, {Primack}, \& {Dekel}}]{ceverino14}
{Ceverino}, D., {Klypin}, A., {Klimek}, E., {et~al.} 2014, ArXiv e-prints

\bibitem[{{Chabrier}(2003)}]{chabrier}
{Chabrier}, G. 2003, \pasp, 115, 763

\bibitem[{{Chang} {et~al.}(2013){Chang}, {van der Wel}, {Rix}, {Holden},
  {Bell}, {McGrath}, {Wuyts}, {H{\"a}u{\ss}ler}, {Barden}, {Faber}, {Mozena},
  {Ferguson}, {Guo}, {Galametz}, {Grogin}, {Kocevski}, {Koekemoer}, {Dekel},
  {Huang}, {Hathi}, \& {Donley}}]{chang13}
{Chang}, Y.-Y., {van der Wel}, A., {Rix}, H.-W., {et~al.} 2013, ArXiv e-prints

\bibitem[{{Cheung} {et~al.}(2012){Cheung}, {Faber}, {Koo}, {Dutton}, {Simard},
  {McGrath}, {Huang}, {Bell}, {Dekel}, {Fang}, {Salim}, {Barro}, {Bundy},
  {Coil}, {Cooper}, {Conselice}, {Davis}, {Dominguez}, {Kassin}, {Kocevski},
  {Koekemoer}, {Lin}, {Lotz}, {Newman}, {Phillips}, {Rosario}, {Weiner}, \&
  {Willmer}}]{cheung12}
{Cheung}, E., {Faber}, S.~M., {Koo}, D.~C., {et~al.} 2012, ArXiv e-prints

\bibitem[{{Courteau} {et~al.}(2014){Courteau}, {Cappellari}, {de Jong},
  {Dutton}, {Emsellem}, {Hoekstra}, {Koopmans}, {Mamon}, {Maraston}, {Treu}, \&
  {Widrow}}]{courteau14}
{Courteau}, S., {Cappellari}, M., {de Jong}, R.~S., {et~al.} 2014, Reviews of
  Modern Physics, 86, 47

\bibitem[{{Daddi} {et~al.}(2010{\natexlab{a}}){Daddi}, {Bournaud}, {Walter},
  {Dannerbauer}, {Carilli}, {Dickinson}, {Elbaz}, {Morrison}, {Riechers},
  {Onodera}, {Salmi}, {Krips}, \& {Stern}}]{daddi10a}
{Daddi}, E., {Bournaud}, F., {Walter}, F., {et~al.} 2010{\natexlab{a}}, \apj,
  713, 686

\bibitem[{{Daddi} {et~al.}(2008){Daddi}, {Dannerbauer}, {Elbaz}, {Dickinson},
  {Morrison}, {Stern}, \& {Ravindranath}}]{daddi08}
{Daddi}, E., {Dannerbauer}, H., {Elbaz}, D., {et~al.} 2008, \apjl, 673, L21

\bibitem[{{Daddi} {et~al.}(2010{\natexlab{b}}){Daddi}, {Elbaz}, {Walter},
  {Bournaud}, {Salmi}, {Carilli}, {Dannerbauer}, {Dickinson}, {Monaco}, \&
  {Riechers}}]{daddi10b}
{Daddi}, E., {Elbaz}, D., {Walter}, F., {et~al.} 2010{\natexlab{b}}, \apjl,
  714, L118

\bibitem[{{De Breuck} {et~al.}(2014){De Breuck}, {Williams}, {Swinbank},
  {Caselli}, {Coppin}, {Davis}, {Maiolino}, {Nagao}, {Smail}, {Walter},
  {Weiss}, \& {Zwaan}}]{debreuck14}
{De Breuck}, C., {Williams}, R.~J., {Swinbank}, M., {et~al.} 2014, ArXiv
  e-prints

\bibitem[{{Dekel} \& {Birnboim}(2006)}]{dekel06}
{Dekel}, A. \& {Birnboim}, Y. 2006, \mnras, 368, 2

\bibitem[{{Dekel} {et~al.}(2009){Dekel}, {Birnboim}, {Engel}, {Freundlich},
  {Goerdt}, {Mumcuoglu}, {Neistein}, {Pichon}, {Teyssier}, \&
  {Zinger}}]{dekel09a}
{Dekel}, A., {Birnboim}, Y., {Engel}, G., {et~al.} 2009, \nat, 457, 451

\bibitem[{{Dekel} \& {Burkert}(2014)}]{dekel13b}
{Dekel}, A. \& {Burkert}, A. 2014, ArXiv e-prints

\bibitem[{{Dekel} {et~al.}(2013){Dekel}, {Zolotov}, {Tweed}, {Cacciato},
  {Ceverino}, \& {Primack}}]{dekel13a}
{Dekel}, A., {Zolotov}, A., {Tweed}, D., {et~al.} 2013, \mnras, 435, 999

\bibitem[{{Donley} {et~al.}(2012){Donley}, {Koekemoer}, {Brusa}, {Capak},
  {Cardamone}, {Civano}, {Ilbert}, {Impey}, {Kartaltepe}, {Miyaji}, {Salvato},
  {Sanders}, {Trump}, \& {Zamorani}}]{donley12}
{Donley}, J.~L., {Koekemoer}, A.~M., {Brusa}, M., {et~al.} 2012, \apj, 748, 142

\bibitem[{{Donley} {et~al.}(2007){Donley}, {Rieke}, {P{\'e}rez-Gonz{\'a}lez},
  {Rigby}, \& {Alonso-Herrero}}]{donley07}
{Donley}, J.~L., {Rieke}, G.~H., {P{\'e}rez-Gonz{\'a}lez}, P.~G., {Rigby},
  J.~R., \& {Alonso-Herrero}, A. 2007, \apj, 660, 167

\bibitem[{{Elbaz} {et~al.}(2011){Elbaz}, {Dickinson}, {Hwang},
  {D{\'{\i}}az-Santos}, {Magdis}, {Magnelli}, {Le Borgne}, {Galliano},
  {Pannella}, {Chanial}, {Armus}, {Charmandaris}, {Daddi}, {Aussel}, {Popesso},
  {Kartaltepe}, {Altieri}, {Valtchanov}, {Coia}, {Dannerbauer}, {Dasyra},
  {Leiton}, {Mazzarella}, {Alexander}, {Buat}, {Burgarella}, {Chary}, {Gilli},
  {Ivison}, {Juneau}, {Le Floc'h}, {Lutz}, {Morrison}, {Mullaney}, {Murphy},
  {Pope}, {Scott}, {Brodwin}, {Calzetti}, {Cesarsky}, {Charlot}, {Dole},
  {Eisenhardt}, {Ferguson}, {F{\"o}rster Schreiber}, {Frayer}, {Giavalisco},
  {Huynh}, {Koekemoer}, {Papovich}, {Reddy}, {Surace}, {Teplitz}, {Yun}, \&
  {Wilson}}]{elbaz11}
{Elbaz}, D., {Dickinson}, M., {Hwang}, H.~S., {et~al.} 2011, \aap, 533, A119

\bibitem[{{Elmegreen} \& {Elmegreen}(2005)}]{elmegreen05}
{Elmegreen}, B.~G. \& {Elmegreen}, D.~M. 2005, \apj, 627, 632

\bibitem[{{Elmegreen} {et~al.}(2004){Elmegreen}, {Elmegreen}, \&
  {Sheets}}]{elmegreen04}
{Elmegreen}, D.~M., {Elmegreen}, B.~G., \& {Sheets}, C.~M. 2004, \apj, 603, 74

\bibitem[{{Engel} {et~al.}(2010){Engel}, {Tacconi}, {Davies}, {Neri}, {Smail},
  {Chapman}, {Genzel}, {Cox}, {Greve}, {Ivison}, {Blain}, {Bertoldi}, \&
  {Omont}}]{engel10}
{Engel}, H., {Tacconi}, L.~J., {Davies}, R.~I., {et~al.} 2010, \apj, 724, 233

\bibitem[{{Epinat} {et~al.}(2012){Epinat}, {Tasca}, {Amram}, {Contini}, {Le
  F{\`e}vre}, {Queyrel}, {Vergani}, {Garilli}, {Kissler-Patig}, {Moultaka},
  {Paioro}, {Tresse}, {Bournaud}, {L{\'o}pez-Sanjuan}, \& {Perret}}]{epinat12}
{Epinat}, B., {Tasca}, L., {Amram}, P., {et~al.} 2012, \aap, 539, A92

\bibitem[{{Erb} {et~al.}(2006){Erb}, {Steidel}, {Shapley}, {Pettini}, {Reddy},
  \& {Adelberger}}]{erb06}
{Erb}, D.~K., {Steidel}, C.~C., {Shapley}, A.~E., {et~al.} 2006, \apj, 647, 128

\bibitem[{{Faber} \& {Jackson}(1976)}]{faber76}
{Faber}, S.~M. \& {Jackson}, R.~E. 1976, \apj, 204, 668

\bibitem[{{F{\"o}rster Schreiber} {et~al.}(2009){F{\"o}rster Schreiber},
  {Genzel}, {Bouch{\'e}}, {Cresci}, {Davies}, {Buschkamp}, {Shapiro},
  {Tacconi}, {Hicks}, {Genel}, {Shapley}, {Erb}, {Steidel}, {Lutz},
  {Eisenhauer}, {Gillessen}, {Sternberg}, {Renzini}, {Cimatti}, {Daddi},
  {Kurk}, {Lilly}, {Kong}, {Lehnert}, {Nesvadba}, {Verma}, {McCracken},
  {Arimoto}, {Mignoli}, \& {Onodera}}]{fs09}
{F{\"o}rster Schreiber}, N.~M., {Genzel}, R., {Bouch{\'e}}, N., {et~al.} 2009,
  \apj, 706, 1364

\bibitem[{{F{\"o}rster Schreiber} {et~al.}(2011){F{\"o}rster Schreiber},
  {Shapley}, {Erb}, {Genzel}, {Steidel}, {Bouch{\'e}}, {Cresci}, \&
  {Davies}}]{fs11}
{F{\"o}rster Schreiber}, N.~M., {Shapley}, A.~E., {Erb}, D.~K., {et~al.} 2011,
  \apj, 731, 65

\bibitem[{{Franx} {et~al.}(2008){Franx}, {van Dokkum}, {Schreiber}, {Wuyts},
  {Labb{\'e}}, \& {Toft}}]{franx08}
{Franx}, M., {van Dokkum}, P.~G., {Schreiber}, N.~M.~F., {et~al.} 2008, \apj,
  688, 770

\bibitem[{{Fu} \& {Stockton}(2009)}]{fu09}
{Fu}, H. \& {Stockton}, A. 2009, \apj, 696, 1693

\bibitem[{{Fumagalli} {et~al.}(2013){Fumagalli}, {Labbe}, {Patel}, {Franx},
  {van Dokkum}, {Brammer}, {da Cunha}, {Forster Schreiber}, {Kriek}, {Quadri},
  {Rix}, {Wake}, {Whitaker}, {Lundgren}, {Marchesini}, {Maseda}, {Momcheva},
  {Nelson}, {Pacifici}, \& {Skelton}}]{fumagalli13}
{Fumagalli}, M., {Labbe}, I., {Patel}, S.~G., {et~al.} 2013, ArXiv e-prints

\bibitem[{{Gallazzi} {et~al.}(2006){Gallazzi}, {Charlot}, {Brinchmann}, \&
  {White}}]{gallazzi06}
{Gallazzi}, A., {Charlot}, S., {Brinchmann}, J., \& {White}, S.~D.~M. 2006,
  \mnras, 370, 1106

\bibitem[{{Genzel} {et~al.}(2008){Genzel}, {Burkert}, {Bouch{\'e}}, {Cresci},
  {F{\"o}rster Schreiber}, {Shapley}, {Shapiro}, {Tacconi}, {Buschkamp},
  {Cimatti}, {Daddi}, {Davies}, {Eisenhauer}, {Erb}, {Genel}, {Gerhard},
  {Hicks}, {Lutz}, {Naab}, {Ott}, {Rabien}, {Renzini}, {Steidel}, {Sternberg},
  \& {Lilly}}]{genzel08}
{Genzel}, R., {Burkert}, A., {Bouch{\'e}}, N., {et~al.} 2008, \apj, 687, 59

\bibitem[{{Genzel} {et~al.}(2012){Genzel}, {Tacconi}, {Combes}, {Bolatto},
  {Neri}, {Sternberg}, {Cooper}, {Bouch{\'e}}, {Bournaud}, {Burkert},
  {Comerford}, {Cox}, {Davis}, {F{\"o}rster Schreiber}, {Garcia-Burillo},
  {Gracia-Carpio}, {Lutz}, {Naab}, {Newman}, {Saintonge}, {Shapiro}, {Shapley},
  \& {Weiner}}]{genzel12}
{Genzel}, R., {Tacconi}, L.~J., {Combes}, F., {et~al.} 2012, \apj, 746, 69

\bibitem[{{Genzel} {et~al.}(2010){Genzel}, {Tacconi}, {Gracia-Carpio},
  {Sternberg}, {Cooper}, {Shapiro}, {Bolatto}, {Bouch{\'e}}, {Bournaud},
  {Burkert}, {Combes}, {Comerford}, {Cox}, {Davis}, {Schreiber},
  {Garcia-Burillo}, {Lutz}, {Naab}, {Neri}, {Omont}, {Shapley}, \&
  {Weiner}}]{genzel10}
{Genzel}, R., {Tacconi}, L.~J., {Gracia-Carpio}, J., {et~al.} 2010, \mnras,
  407, 2091

\bibitem[{{Gilli} {et~al.}(2014){Gilli}, {Norman}, {Vignali}, {Vanzella},
  {Calura}, {Pozzi}, {Massardi}, {Mignano}, {Casasola}, {Daddi}, {Elbaz},
  {Dickinson}, {Iwasawa}, {Maiolino}, {Brusa}, {Vito}, {Fritz}, {Feltre},
  {Cresci}, {Mignoli}, {Comastri}, \& {Zamorani}}]{gilli14}
{Gilli}, R., {Norman}, C., {Vignali}, C., {et~al.} 2014, \aap, 562, A67

\bibitem[{{Grogin} {et~al.}(2011){Grogin}, {Kocevski}, {Faber}, {Ferguson},
  {Koekemoer}, {Riess}, {Acquaviva}, {Alexander}, {Almaini}, {Ashby}, {Barden},
  {Bell}, {Bournaud}, {Brown}, {Caputi}, {Casertano}, {Cassata}, {Castellano},
  {Challis}, {Chary}, {Cheung}, {Cirasuolo}, {Conselice}, {Roshan Cooray},
  {Croton}, {Daddi}, {Dahlen}, {Dav{\'e}}, {de Mello}, {Dekel}, {Dickinson},
  {Dolch}, {Donley}, {Dunlop}, {Dutton}, {Elbaz}, {Fazio}, {Filippenko},
  {Finkelstein}, {Fontana}, {Gardner}, {Garnavich}, {Gawiser}, {Giavalisco},
  {Grazian}, {Guo}, {Hathi}, {H{\"a}ussler}, {Hopkins}, {Huang}, {Huang},
  {Jha}, {Kartaltepe}, {Kirshner}, {Koo}, {Lai}, {Lee}, {Li}, {Lotz}, {Lucas},
  {Madau}, {McCarthy}, {McGrath}, {McIntosh}, {McLure}, {Mobasher},
  {Moustakas}, {Mozena}, {Nandra}, {Newman}, {Niemi}, {Noeske}, {Papovich},
  {Pentericci}, {Pope}, {Primack}, {Rajan}, {Ravindranath}, {Reddy}, {Renzini},
  {Rix}, {Robaina}, {Rodney}, {Rosario}, {Rosati}, {Salimbeni}, {Scarlata},
  {Siana}, {Simard}, {Smidt}, {Somerville}, {Spinrad}, {Straughn}, {Strolger},
  {Telford}, {Teplitz}, {Trump}, {van der Wel}, {Villforth}, {Wechsler},
  {Weiner}, {Wiklind}, {Wild}, {Wilson}, {Wuyts}, {Yan}, \& {Yun}}]{candelsgro}
{Grogin}, N.~A., {Kocevski}, D.~D., {Faber}, S.~M., {et~al.} 2011, \apjs, 197,
  35

\bibitem[{{Guo} {et~al.}(2013){Guo}, {Ferguson}, {Giavalisco}, {Barro},
  {Willner}, {Ashby}, {Dahlen}, {Donley}, {Faber}, {Fontana}, {Galametz},
  {Grazian}, {Huang}, {Kocevski}, {Koekemoer}, {Koo}, {McGrath}, {Peth},
  {Salvato}, {Wuyts}, {Castellano}, {Cooray}, {Dickinson}, {Dunlop}, {Fazio},
  {Gardner}, {Gawiser}, {Grogin}, {Hathi}, {Hsu}, {Lee}, {Lucas}, {Mobasher},
  {Nandra}, {Newman}, \& {van der Wel}}]{guo13}
{Guo}, Y., {Ferguson}, H.~C., {Giavalisco}, M., {et~al.} 2013, \apjs, 207, 24

\bibitem[{{Guo} {et~al.}(2012){Guo}, {Giavalisco}, {Cassata}, {Ferguson},
  {Williams}, {Dickinson}, {Koekemoer}, {Grogin}, {Chary}, {Messias}, {Tundo},
  {Lin}, {Lee}, {Salimbeni}, {Fontana}, {Grazian}, {Kocevski}, {Lee},
  {Villanueva}, \& {van der Wel}}]{guo12}
{Guo}, Y., {Giavalisco}, M., {Cassata}, P., {et~al.} 2012, \apj, 749, 149

\bibitem[{{Hernquist} \& {Mihos}(1995)}]{hernquist95}
{Hernquist}, L. \& {Mihos}, J.~C. 1995, \apj, 448, 41

\bibitem[{{Holden} {et~al.}(2014){Holden}, {Oesch}, {Gonzalez}, {Illingworth},
  {Labbe}, {Bouwens}, {Franx}, {van Dokkum}, \& {Spitler}}]{holden14}
{Holden}, B.~P., {Oesch}, P.~A., {Gonzalez}, V.~G., {et~al.} 2014, ArXiv
  e-prints

\bibitem[{{Hopkins} {et~al.}(2006){Hopkins}, {Hernquist}, {Cox}, {Di Matteo},
  {Robertson}, \& {Springel}}]{hopkins06}
{Hopkins}, P.~F., {Hernquist}, L., {Cox}, T.~J., {et~al.} 2006, \apjs, 163, 1

\bibitem[{{Kassin} {et~al.}(2007){Kassin}, {Weiner}, {Faber}, {Koo}, {Lotz},
  {Diemand}, {Harker}, {Bundy}, {Metevier}, {Phillips}, {Cooper}, {Croton},
  {Konidaris}, {Noeske}, \& {Willmer}}]{kassin07}
{Kassin}, S.~A., {Weiner}, B.~J., {Faber}, S.~M., {et~al.} 2007, \apjl, 660,
  L35

\bibitem[{{Kennicutt}(1998)}]{ken98}
{Kennicutt}, Jr., R.~C. 1998, \araa, 36, 189

\bibitem[{{Kere{\v s}} {et~al.}(2005){Kere{\v s}}, {Katz}, {Weinberg}, \&
  {Dav{\'e}}}]{keres05}
{Kere{\v s}}, D., {Katz}, N., {Weinberg}, D.~H., \& {Dav{\'e}}, R. 2005,
  \mnras, 363, 2

\bibitem[{{Kirkpatrick} {et~al.}(2013){Kirkpatrick}, {Pope}, {Charmandaris},
  {Daddi}, {Elbaz}, {Hwang}, {Pannella}, {Scott}, {Altieri}, {Aussel}, {Coia},
  {Dannerbauer}, {Dasyra}, {Dickinson}, {Kartaltepe}, {Leiton}, {Magdis},
  {Magnelli}, {Popesso}, \& {Valtchanov}}]{kirkpatric13}
{Kirkpatrick}, A., {Pope}, A., {Charmandaris}, V., {et~al.} 2013, \apj, 763,
  123

\bibitem[{{Koekemoer} {et~al.}(2011){Koekemoer}, {Faber}, {Ferguson}, {Grogin},
  {Kocevski}, {Koo}, {Lai}, {Lotz}, {Lucas}, {McGrath}, {Ogaz}, {Rajan},
  {Riess}, {Rodney}, {Strolger}, {Casertano}, {Castellano}, {Dahlen},
  {Dickinson}, {Dolch}, {Fontana}, {Giavalisco}, {Grazian}, {Guo}, {Hathi},
  {Huang}, {van der Wel}, {Yan}, {Acquaviva}, {Alexander}, {Almaini}, {Ashby},
  {Barden}, {Bell}, {Bournaud}, {Brown}, {Caputi}, {Cassata}, {Challis},
  {Chary}, {Cheung}, {Cirasuolo}, {Conselice}, {Roshan Cooray}, {Croton},
  {Daddi}, {Dav{\'e}}, {de Mello}, {de Ravel}, {Dekel}, {Donley}, {Dunlop},
  {Dutton}, {Elbaz}, {Fazio}, {Filippenko}, {Finkelstein}, {Frazer}, {Gardner},
  {Garnavich}, {Gawiser}, {Gruetzbauch}, {Hartley}, {H{\"a}ussler},
  {Herrington}, {Hopkins}, {Huang}, {Jha}, {Johnson}, {Kartaltepe},
  {Khostovan}, {Kirshner}, {Lani}, {Lee}, {Li}, {Madau}, {McCarthy},
  {McIntosh}, {McLure}, {McPartland}, {Mobasher}, {Moreira}, {Mortlock},
  {Moustakas}, {Mozena}, {Nandra}, {Newman}, {Nielsen}, {Niemi}, {Noeske},
  {Papovich}, {Pentericci}, {Pope}, {Primack}, {Ravindranath}, {Reddy},
  {Renzini}, {Rix}, {Robaina}, {Rosario}, {Rosati}, {Salimbeni}, {Scarlata},
  {Siana}, {Simard}, {Smidt}, {Snyder}, {Somerville}, {Spinrad}, {Straughn},
  {Telford}, {Teplitz}, {Trump}, {Vargas}, {Villforth}, {Wagner}, {Wandro},
  {Wechsler}, {Weiner}, {Wiklind}, {Wild}, {Wilson}, {Wuyts}, \&
  {Yun}}]{candelskoe}
{Koekemoer}, A.~M., {Faber}, S.~M., {Ferguson}, H.~C., {et~al.} 2011, \apjs,
  197, 36

\bibitem[{{Kravtsov} {et~al.}(1997){Kravtsov}, {Klypin}, \& {Khokhlov}}]{arm}
{Kravtsov}, A.~V., {Klypin}, A.~A., \& {Khokhlov}, A.~M. 1997, \apjs, 111, 73

\bibitem[{{Kriek} {et~al.}(2009){Kriek}, {van Dokkum}, {Labb{\'e}}, {Franx},
  {Illingworth}, {Marchesini}, \& {Quadri}}]{fast}
{Kriek}, M., {van Dokkum}, P.~G., {Labb{\'e}}, I., {et~al.} 2009, \apj, 700,
  221

\bibitem[{{Krist}(1995)}]{tinytim}
{Krist}, J. 1995, in Astronomical Society of the Pacific Conference Series,
  Vol.~77, Astronomical Data Analysis Software and Systems IV, ed. R.~A.
  {Shaw}, H.~E. {Payne}, \& J.~J.~E. {Hayes}, 349

\bibitem[{{Laird} {et~al.}(2010){Laird}, {Nandra}, {Pope}, \&
  {Scott}}]{laird10}
{Laird}, E.~S., {Nandra}, K., {Pope}, A., \& {Scott}, D. 2010, \mnras, 401,
  2763

\bibitem[{{Law} {et~al.}(2007){Law}, {Steidel}, {Erb}, {Larkin}, {Pettini},
  {Shapley}, \& {Wright}}]{law07}
{Law}, D.~R., {Steidel}, C.~C., {Erb}, D.~K., {et~al.} 2007, \apj, 669, 929

\bibitem[{{Law} {et~al.}(2009){Law}, {Steidel}, {Erb}, {Larkin}, {Pettini},
  {Shapley}, \& {Wright}}]{law09}
---. 2009, \apj, 697, 2057

\bibitem[{{Law} {et~al.}(2012){Law}, {Steidel}, {Shapley}, {Nagy}, {Reddy}, \&
  {Erb}}]{law12a}
{Law}, D.~R., {Steidel}, C.~C., {Shapley}, A.~E., {et~al.} 2012, \apj, 745, 85

\bibitem[{{Magdis} {et~al.}(2012){Magdis}, {Daddi}, {B{\'e}thermin}, {Sargent},
  {Elbaz}, {Pannella}, {Dickinson}, {Dannerbauer}, {da Cunha}, {Walter},
  {Rigopoulou}, {Charmandaris}, {Hwang}, \& {Kartaltepe}}]{magdis12}
{Magdis}, G.~E., {Daddi}, E., {B{\'e}thermin}, M., {et~al.} 2012, \apj, 760, 6

\bibitem[{{Magnelli} {et~al.}(2012){Magnelli}, {Lutz}, {Santini}, {Saintonge},
  {Berta}, {Albrecht}, {Altieri}, {Andreani}, {Aussel}, {Bertoldi},
  {B{\'e}thermin}, {Bongiovanni}, {Capak}, {Chapman}, {Cepa}, {Cimatti},
  {Cooray}, {Daddi}, {Danielson}, {Dannerbauer}, {Dunlop}, {Elbaz}, {Farrah},
  {F{\"o}rster Schreiber}, {Genzel}, {Hwang}, {Ibar}, {Ivison}, {Le Floc'h},
  {Magdis}, {Maiolino}, {Nordon}, {Oliver}, {P{\'e}rez Garc{\'{\i}}a},
  {Poglitsch}, {Popesso}, {Pozzi}, {Riguccini}, {Rodighiero}, {Rosario},
  {Roseboom}, {Salvato}, {Sanchez-Portal}, {Scott}, {Smail}, {Sturm},
  {Swinbank}, {Tacconi}, {Valtchanov}, {Wang}, \& {Wuyts}}]{magnelli12a}
{Magnelli}, B., {Lutz}, D., {Santini}, P., {et~al.} 2012, \aap, 539, A155

\bibitem[{{Magnelli} {et~al.}(2013){Magnelli}, {Popesso}, {Berta}, {Pozzi},
  {Elbaz}, {Lutz}, {Dickinson}, {Altieri}, {Andreani}, {Aussel},
  {B{\'e}thermin}, {Bongiovanni}, {Cepa}, {Charmandaris}, {Chary}, {Cimatti},
  {Daddi}, {F{\"o}rster Schreiber}, {Genzel}, {Gruppioni}, {Harwit}, {Hwang},
  {Ivison}, {Magdis}, {Maiolino}, {Murphy}, {Nordon}, {Pannella}, {P{\'e}rez
  Garc{\'{\i}}a}, {Poglitsch}, {Rosario}, {Sanchez-Portal}, {Santini}, {Scott},
  {Sturm}, {Tacconi}, \& {Valtchanov}}]{magnelli13}
{Magnelli}, B., {Popesso}, P., {Berta}, S., {et~al.} 2013, \aap, 553, A132

\bibitem[{{Maseda} {et~al.}(2013){Maseda}, {van der Wel}, {da Cunha}, {Rix},
  {Pacifici}, {Momcheva}, {Brammer}, {Franx}, {van Dokkum}, {Bell},
  {Fumagalli}, {Grogin}, {Kocevski}, {Koekemoer}, {Lundgren}, {Marchesini},
  {Nelson}, {Patel}, {Skelton}, {Straughn}, {Trump}, {Weiner}, {Whitaker}, \&
  {Wuyts}}]{maseda13}
{Maseda}, M.~V., {van der Wel}, A., {da Cunha}, E., {et~al.} 2013, \apjl, 778,
  L22

\bibitem[{{Masters} \& {Capak}(2011)}]{specpro}
{Masters}, D. \& {Capak}, P. 2011, \pasp, 123, 638

\bibitem[{{Masters} {et~al.}(2014){Masters}, {McCarthy}, {Siana}, {Malkan},
  {Mobasher}, {Atek}, {Henry}, {Martin}, {Rafelski}, {Hathi}, {Scarlata},
  {Ross}, {Bunker}, {Blanc}, {Bedregal}, {Dominguez}, {Colbert}, {Teplitz}, \&
  {Dressler}}]{masters14}
{Masters}, D., {McCarthy}, P., {Siana}, B., {et~al.} 2014, ArXiv e-prints

\bibitem[{{McLean} {et~al.}(2010){McLean}, {Steidel}, {Epps}, {Matthews},
  {Adkins}, {Konidaris}, {Weber}, {Aliado}, {Brims}, {Canfield}, {Cromer},
  {Fucik}, {Kulas}, {Mace}, {Magnone}, {Rodriguez}, {Wang}, \&
  {Weiss}}]{mosfire1}
{McLean}, I.~S., {Steidel}, C.~C., {Epps}, H., {et~al.} 2010, in Society of
  Photo-Optical Instrumentation Engineers (SPIE) Conference Series, Vol. 7735,
  Society of Photo-Optical Instrumentation Engineers (SPIE) Conference Series

\bibitem[{{McLean} {et~al.}(2012){McLean}, {Steidel}, {Epps}, {Konidaris},
  {Matthews}, {Adkins}, {Aliado}, {Brims}, {Canfield}, {Cromer}, {Fucik},
  {Kulas}, {Mace}, {Magnone}, {Rodriguez}, {Rudie}, {Trainor}, {Wang}, {Weber},
  \& {Weiss}}]{mosfire2}
{McLean}, I.~S., {Steidel}, C.~C., {Epps}, H.~W., {et~al.} 2012, in Society of
  Photo-Optical Instrumentation Engineers (SPIE) Conference Series, Vol. 8446,
  Society of Photo-Optical Instrumentation Engineers (SPIE) Conference Series

\bibitem[{{Micha{\l}owski} {et~al.}(2012){Micha{\l}owski}, {Dunlop},
  {Cirasuolo}, {Hjorth}, {Hayward}, \& {Watson}}]{michalowski12}
{Micha{\l}owski}, M.~J., {Dunlop}, J.~S., {Cirasuolo}, M., {et~al.} 2012, \aap,
  541, A85

\bibitem[{{Muzzin} {et~al.}(2013){Muzzin}, {Marchesini}, {Stefanon}, {Franx},
  {McCracken}, {Milvang-Jensen}, {Dunlop}, {Fynbo}, {Le Fevre}, {Brammer}, \&
  {Labbe}}]{muzzin13smf}
{Muzzin}, A., {Marchesini}, D., {Stefanon}, M., {et~al.} 2013, ArXiv e-prints

\bibitem[{{Naab} {et~al.}(2007){Naab}, {Johansson}, {Ostriker}, \&
  {Efstathiou}}]{naab07}
{Naab}, T., {Johansson}, P.~H., {Ostriker}, J.~P., \& {Efstathiou}, G. 2007,
  \apj, 658, 710

\bibitem[{{Narayanan} {et~al.}(2012){Narayanan}, {Krumholz}, {Ostriker}, \&
  {Hernquist}}]{narayanan12}
{Narayanan}, D., {Krumholz}, M.~R., {Ostriker}, E.~C., \& {Hernquist}, L. 2012,
  \mnras, 421, 3127

\bibitem[{{Newman} {et~al.}(2013{\natexlab{a}}){Newman}, {Ellis}, {Andreon},
  {Treu}, {Raichoor}, \& {Trinchieri}}]{newman13}
{Newman}, A.~B., {Ellis}, R.~S., {Andreon}, S., {et~al.} 2013{\natexlab{a}},
  ArXiv e-prints

\bibitem[{{Newman} {et~al.}(2012){Newman}, {Ellis}, {Bundy}, \&
  {Treu}}]{newman12}
{Newman}, A.~B., {Ellis}, R.~S., {Bundy}, K., \& {Treu}, T. 2012, \apj, 746,
  162

\bibitem[{{Newman} {et~al.}(2013{\natexlab{b}}){Newman}, {Genzel}, {F{\"o}rster
  Schreiber}, {Shapiro Griffin}, {Mancini}, {Lilly}, {Renzini}, {Bouch{\'e}},
  {Burkert}, {Buschkamp}, {Carollo}, {Cresci}, {Davies}, {Eisenhauer}, {Genel},
  {Hicks}, {Kurk}, {Lutz}, {Naab}, {Peng}, {Sternberg}, {Tacconi}, {Wuyts},
  {Zamorani}, \& {Vergani}}]{snewman13}
{Newman}, S.~F., {Genzel}, R., {F{\"o}rster Schreiber}, N.~M., {et~al.}
  2013{\natexlab{b}}, \apj, 767, 104

\bibitem[{{Patel} {et~al.}(2012){Patel}, {van Dokkum}, {Franx}, {Quadri},
  {Muzzin}, {Marchesini}, {Williams}, {Holden}, \& {Stefanon}}]{patel13}
{Patel}, S.~G., {van Dokkum}, P.~G., {Franx}, M., {et~al.} 2012, ArXiv e-prints

\bibitem[{{Peng} {et~al.}(2002){Peng}, {Ho}, {Impey}, \& {Rix}}]{galfit}
{Peng}, C.~Y., {Ho}, L.~C., {Impey}, C.~D., \& {Rix}, H.-W. 2002, \aj, 124, 266

\bibitem[{{P{\'e}rez-Gonz{\'a}lez} {et~al.}(2008){P{\'e}rez-Gonz{\'a}lez},
  {Trujillo}, {Barro}, {Gallego}, {Zamorano}, \& {Conselice}}]{pg08b}
{P{\'e}rez-Gonz{\'a}lez}, P.~G., {Trujillo}, I., {Barro}, G., {et~al.} 2008,
  \apj, 687, 50

\bibitem[{{Rangel} {et~al.}(2014){Rangel}, {Nandra}, {Barro}, {Brightman},
  {Hsu}, {Salvato}, {Koekemoer}, {Brusa}, {Laird}, {Trump}, {Croton}, {Koo},
  {Kocevski}, {Donley}, {Hathi}, {Peth}, {Faber}, {Mozena}, {Grogin},
  {Ferguson}, \& {Lai}}]{rangel14}
{Rangel}, C., {Nandra}, K., {Barro}, G., {et~al.} 2014, \mnras, 440, 3630

\bibitem[{{Rodighiero} {et~al.}(2011){Rodighiero}, {Daddi}, {Baronchelli},
  {Cimatti}, {Renzini}, {Aussel}, {Popesso}, {Lutz}, {Andreani}, {Berta},
  {Cava}, {Elbaz}, {Feltre}, {Fontana}, {F{\"o}rster Schreiber},
  {Franceschini}, {Genzel}, {Grazian}, {Gruppioni}, {Ilbert}, {Le Floch},
  {Magdis}, {Magliocchetti}, {Magnelli}, {Maiolino}, {McCracken}, {Nordon},
  {Poglitsch}, {Santini}, {Pozzi}, {Riguccini}, {Tacconi}, {Wuyts}, \&
  {Zamorani}}]{rodi11}
{Rodighiero}, G., {Daddi}, E., {Baronchelli}, I., {et~al.} 2011, \apjl, 739,
  L40

\bibitem[{{Shapley} {et~al.}(2004){Shapley}, {Erb}, {Pettini}, {Steidel}, \&
  {Adelberger}}]{shapley04}
{Shapley}, A.~E., {Erb}, D.~K., {Pettini}, M., {Steidel}, C.~C., \&
  {Adelberger}, K.~L. 2004, \apj, 612, 108

\bibitem[{{Solomon} \& {Vanden Bout}(2005)}]{Solomon05}
{Solomon}, P.~M. \& {Vanden Bout}, P.~A. 2005, \araa, 43, 677

\bibitem[{{Stefanon} {et~al.}(2013){Stefanon}, {Marchesini}, {Rudnick},
  {Brammer}, \& {Whitaker}}]{stefanon13}
{Stefanon}, M., {Marchesini}, D., {Rudnick}, G.~H., {Brammer}, G.~B., \&
  {Whitaker}, K.~E. 2013, \apj, 768, 92

\bibitem[{{Steidel} {et~al.}(2014){Steidel}, {Rudie}, {Strom}, {Pettini},
  {Reddy}, {Shapley}, {Trainor}, {Erb}, {Turner}, {Konidaris}, {Kulas}, {Mace},
  {Matthews}, \& {McLean}}]{steidel14}
{Steidel}, C.~C., {Rudie}, G.~C., {Strom}, A.~L., {et~al.} 2014, ArXiv e-prints

\bibitem[{{Swinbank} {et~al.}(2012{\natexlab{a}}){Swinbank}, {Smail}, {Sobral},
  {Theuns}, {Best}, \& {Geach}}]{swinbank12c}
{Swinbank}, A.~M., {Smail}, I., {Sobral}, D., {et~al.} 2012{\natexlab{a}},
  \apj, 760, 130

\bibitem[{{Swinbank} {et~al.}(2012{\natexlab{b}}){Swinbank}, {Sobral}, {Smail},
  {Geach}, {Best}, {McCarthy}, {Crain}, \& {Theuns}}]{swinbank12b}
{Swinbank}, A.~M., {Sobral}, D., {Smail}, I., {et~al.} 2012{\natexlab{b}},
  \mnras, 426, 935

\bibitem[{{Szomoru} {et~al.}(2012){Szomoru}, {Franx}, \& {van Dokkum}}]{szo12}
{Szomoru}, D., {Franx}, M., \& {van Dokkum}, P.~G. 2012, \apj, 749, 121

\bibitem[{{Tacconi} {et~al.}(2010){Tacconi}, {Genzel}, {Neri}, {Cox}, {Cooper},
  {Shapiro}, {Bolatto}, {Bouch{\'e}}, {Bournaud}, {Burkert}, {Combes},
  {Comerford}, {Davis}, {Schreiber}, {Garcia-Burillo}, {Gracia-Carpio}, {Lutz},
  {Naab}, {Omont}, {Shapley}, {Sternberg}, \& {Weiner}}]{tacconi10}
{Tacconi}, L.~J., {Genzel}, R., {Neri}, R., {et~al.} 2010, \nat, 463, 781

\bibitem[{{Tacconi} {et~al.}(2008){Tacconi}, {Genzel}, {Smail}, {Neri},
  {Chapman}, {Ivison}, {Blain}, {Cox}, {Omont}, {Bertoldi}, {Greve},
  {F{\"o}rster Schreiber}, {Genel}, {Lutz}, {Swinbank}, {Shapley}, {Erb},
  {Cimatti}, {Daddi}, \& {Baker}}]{tacconi08}
{Tacconi}, L.~J., {Genzel}, R., {Smail}, I., {et~al.} 2008, \apj, 680, 246

\bibitem[{{Tacconi} {et~al.}(2013){Tacconi}, {Neri}, {Genzel}, {Combes},
  {Bolatto}, {Cooper}, {Wuyts}, {Bournaud}, {Burkert}, {Comerford}, {Cox},
  {Davis}, {F{\"o}rster Schreiber}, {Garc{\'{\i}}a-Burillo}, {Gracia-Carpio},
  {Lutz}, {Naab}, {Newman}, {Omont}, {Saintonge}, {Shapiro Griffin}, {Shapley},
  {Sternberg}, \& {Weiner}}]{tacconi13}
{Tacconi}, L.~J., {Neri}, R., {Genzel}, R., {et~al.} 2013, \apj, 768, 74

\bibitem[{{Toft} {et~al.}(2012){Toft}, {Gallazzi}, {Zirm}, {Wold}, {Zibetti},
  {Grillo}, \& {Man}}]{toft12}
{Toft}, S., {Gallazzi}, A., {Zirm}, A., {et~al.} 2012, \apj, 754, 3

\bibitem[{{Toft} {et~al.}(2014){Toft}, {Smol{\v c}i{\'c}}, {Magnelli}, {Karim},
  {Zirm}, {Michalowski}, {Capak}, {Sheth}, {Schawinski}, {Krogager}, {Wuyts},
  {Sanders}, {Man}, {Lutz}, {Staguhn}, {Berta}, {Mccracken}, {Krpan}, \&
  {Riechers}}]{toft14}
{Toft}, S., {Smol{\v c}i{\'c}}, V., {Magnelli}, B., {et~al.} 2014, \apj, 782,
  68

\bibitem[{{Trujillo} {et~al.}(2007){Trujillo}, {Conselice}, {Bundy}, {Cooper},
  {Eisenhardt}, \& {Ellis}}]{trujillo07}
{Trujillo}, I., {Conselice}, C.~J., {Bundy}, K., {et~al.} 2007, \mnras, 382,
  109

\bibitem[{{Trump} {et~al.}(2013){Trump}, {Konidaris}, {Barro}, {Koo},
  {Kocevski}, {Juneau}, {Weiner}, {Faber}, {McLean}, {Yan},
  {P{\'e}rez-Gonz{\'a}lez}, \& {Villar}}]{trump13}
{Trump}, J.~R., {Konidaris}, N.~P., {Barro}, G., {et~al.} 2013, \apjl, 763, L6

\bibitem[{{Trump} {et~al.}(2011){Trump}, {Weiner}, {Scarlata}, {Kocevski},
  {Bell}, {McGrath}, {Koo}, {Faber}, {Laird}, {Mozena}, {Rangel}, {Yan},
  {Yesuf}, {Atek}, {Dickinson}, {Donley}, {Dunlop}, {Ferguson}, {Finkelstein},
  {Grogin}, {Hathi}, {Juneau}, {Kartaltepe}, {Koekemoer}, {Nandra}, {Newman},
  {Rodney}, {Straughn}, \& {Teplitz}}]{trump11b}
{Trump}, J.~R., {Weiner}, B.~J., {Scarlata}, C., {et~al.} 2011, \apj, 743, 144

\bibitem[{{Tully} \& {Fisher}(1977)}]{tully77}
{Tully}, R.~B. \& {Fisher}, J.~R. 1977, \aap, 54, 661

\bibitem[{{Utomo} {et~al.}(2014){Utomo}, {Kriek}, {Labbe}, {Conroy}, \&
  {Fumagalli}}]{utomo14}
{Utomo}, D., {Kriek}, M., {Labbe}, I., {Conroy}, C., \& {Fumagalli}, M. 2014,
  ArXiv e-prints

\bibitem[{{van de Sande} {et~al.}(2013){van de Sande}, {Kriek}, {Franx}, {van
  Dokkum}, {Bezanson}, {Bouwens}, {Quadri}, {Rix}, \& {Skelton}}]{vandesande13}
{van de Sande}, J., {Kriek}, M., {Franx}, M., {et~al.} 2013, \apj, 771, 85

\bibitem[{{van der Wel} {et~al.}(2012){van der Wel}, {Bell}, {H{\"a}ussler},
  {McGrath}, {Chang}, {Guo}, {McIntosh}, {Rix}, {Barden}, {Cheung}, {Faber},
  {Ferguson}, {Galametz}, {Grogin}, {Hartley}, {Kartaltepe}, {Kocevski},
  {Koekemoer}, {Lotz}, {Mozena}, {Peth}, \& {Peng}}]{vdw12}
{van der Wel}, A., {Bell}, E.~F., {H{\"a}ussler}, B., {et~al.} 2012, \apjs,
  203, 24

\bibitem[{{van der Wel} {et~al.}(2014){van der Wel}, {Franx}, {van Dokkum},
  {Skelton}, {Momcheva}, {Whitaker}, {Brammer}, {Bell}, {Rix}, {Wuyts},
  {Ferguson}, {Holden}, {Barro}, {Koekemoer}, {Chang}, {McGrath}, {Haussler},
  {Dekel}, {Behroozi}, {Fumagalli}, {Leja}, {Lundgren}, {Maseda}, {Nelson},
  {Wake}, {Patel}, {Labbe}, {Faber}, {Grogin}, \& {Kocevski}}]{vdw14}
{van der Wel}, A., {Franx}, M., {van Dokkum}, P.~G., {et~al.} 2014, ArXiv
  e-prints

\bibitem[{{van der Wel} {et~al.}(2011){van der Wel}, {Rix}, {Wuyts}, {McGrath},
  {Koekemoer}, {Bell}, {Holden}, {Robaina}, \& {McIntosh}}]{vdw11a}
{van der Wel}, A., {Rix}, H.-W., {Wuyts}, S., {et~al.} 2011, \apj, 730, 38

\bibitem[{{van Dokkum} \& {Brammer}(2010)}]{dokkum10b}
{van Dokkum}, P.~G. \& {Brammer}, G. 2010, \apjl, 718, L73

\bibitem[{{van Dokkum} {et~al.}(2008){van Dokkum}, {Franx}, {Kriek}, {Holden},
  {Illingworth}, {Magee}, {Bouwens}, {Marchesini}, {Quadri}, {Rudnick},
  {Taylor}, \& {Toft}}]{dokku08}
{van Dokkum}, P.~G., {Franx}, M., {Kriek}, M., {et~al.} 2008, \apjl, 677, L5

\bibitem[{{Wake} {et~al.}(2012){Wake}, {van Dokkum}, \& {Franx}}]{wake12}
{Wake}, D.~A., {van Dokkum}, P.~G., \& {Franx}, M. 2012, \apjl, 751, L44

\bibitem[{{Weiner} {et~al.}(2006){Weiner}, {Willmer}, {Faber}, {Melbourne},
  {Kassin}, {Phillips}, {Harker}, {Metevier}, {Vogt}, \& {Koo}}]{weiner06}
{Weiner}, B.~J., {Willmer}, C.~N.~A., {Faber}, S.~M., {et~al.} 2006, \apj, 653,
  1027

\bibitem[{{Whitaker} {et~al.}(2011){Whitaker}, {Labb{\'e}}, {van Dokkum},
  {Brammer}, {Kriek}, {Marchesini}, {Quadri}, {Franx}, {Muzzin}, {Williams},
  {Bezanson}, {Illingworth}, {Lee}, {Lundgren}, {Nelson}, {Rudnick}, {Tal}, \&
  {Wake}}]{whitaker11}
{Whitaker}, K.~E., {Labb{\'e}}, I., {van Dokkum}, P.~G., {et~al.} 2011, \apj,
  735, 86

\bibitem[{{Whitaker} {et~al.}(2013){Whitaker}, {van Dokkum}, {Brammer},
  {Momcheva}, {Skelton}, {Franx}, {Kriek}, {Labbe}, {Fumagalli}, {Lundgren},
  {Nelson}, {Patel}, \& {Rix}}]{whitaker13}
{Whitaker}, K.~E., {van Dokkum}, P.~G., {Brammer}, G., {et~al.} 2013, ArXiv
  e-prints

\bibitem[{{Williams} {et~al.}(2014){Williams}, {Maiolino}, {Santini},
  {Marconi}, {Cresci}, {Mannucci}, \& {Lutz}}]{rwilliams14}
{Williams}, R.~J., {Maiolino}, R., {Santini}, P., {et~al.} 2014, ArXiv e-prints

\bibitem[{{Wuyts} {et~al.}(2010){Wuyts}, {Cox}, {Hayward}, {Franx},
  {Hernquist}, {Hopkins}, {Jonsson}, \& {van Dokkum}}]{wuyts10}
{Wuyts}, S., {Cox}, T.~J., {Hayward}, C.~C., {et~al.} 2010, \apj, 722, 1666

\bibitem[{{Wuyts} {et~al.}(2012){Wuyts}, {F{\"o}rster Schreiber}, {Genzel},
  {Guo}, {Barro}, {Bell}, {Dekel}, {Faber}, {Ferguson}, {Giavalisco}, {Grogin},
  {Hathi}, {Huang}, {Kocevski}, {Koekemoer}, {Koo}, {Lotz}, {Lutz}, {McGrath},
  {Newman}, {Rosario}, {Saintonge}, {Tacconi}, {Weiner}, \& {van der
  Wel}}]{wuyts12}
{Wuyts}, S., {F{\"o}rster Schreiber}, N.~M., {Genzel}, R., {et~al.} 2012, \apj,
  753, 114

\bibitem[{{Wuyts} {et~al.}(2011{\natexlab{a}}){Wuyts}, {F{\"o}rster Schreiber},
  {Lutz}, {Nordon}, {Berta}, {Altieri}, {Andreani}, {Aussel}, {Bongiovanni},
  {Cepa}, {Cimatti}, {Daddi}, {Elbaz}, {Genzel}, {Koekemoer}, {Magnelli},
  {Maiolino}, {McGrath}, {P{\'e}rez Garc{\'{\i}}a}, {Poglitsch}, {Popesso},
  {Pozzi}, {Sanchez-Portal}, {Sturm}, {Tacconi}, \& {Valtchanov}}]{wuyts11a}
{Wuyts}, S., {F{\"o}rster Schreiber}, N.~M., {Lutz}, D., {et~al.}
  2011{\natexlab{a}}, \apj, 738, 106

\bibitem[{{Wuyts} {et~al.}(2011{\natexlab{b}}){Wuyts}, {F{\"o}rster Schreiber},
  {van der Wel}, {Magnelli}, {Guo}, {Genzel}, {Lutz}, {Aussel}, {Barro},
  {Berta}, {Cava}, {Graci{\'a}-Carpio}, {Hathi}, {Huang}, {Kocevski},
  {Koekemoer}, {Lee}, {Le Floc'h}, {McGrath}, {Nordon}, {Popesso}, {Pozzi},
  {Riguccini}, {Rodighiero}, {Saintonge}, \& {Tacconi}}]{wuyts11b}
{Wuyts}, S., {F{\"o}rster Schreiber}, N.~M., {van der Wel}, A., {et~al.}
  2011{\natexlab{b}}, \apj, 742, 96

\bibitem[{{Xue} {et~al.}(2011){Xue}, {Luo}, {Brandt}, {Bauer}, {Lehmer},
  {Broos}, {Schneider}, {Alexander}, {Brusa}, {Comastri}, {Fabian}, {Gilli},
  {Hasinger}, {Hornschemeier}, {Koekemoer}, {Liu}, {Mainieri}, {Paolillo},
  {Rafferty}, {Rosati}, {Shemmer}, {Silverman}, {Smail}, {Tozzi}, \&
  {Vignali}}]{chandra4m}
{Xue}, Y.~Q., {Luo}, B., {Brandt}, W.~N., {et~al.} 2011, \apjs, 195, 10

\end{thebibliography}
\clearpage
\label{lastpage}
\end{document}